\documentclass[reprint,aps,superscriptaddress,amsmath,longbibliography]{revtex4-2}
\usepackage{graphicx,txfonts,dcolumn,multirow,color}
\usepackage[colorlinks,linkcolor=blue,citecolor=blue,anchorcolor=blue,urlcolor=blue]{hyperref}

\begin{document}

\title{Self-similar Dynamics in Percolation and Sandpile}

\author{Mingzhong Lu}
\affiliation{Department of Modern Physics, University of Science and Technology of China, Hefei, Anhui 230026, China}

\author{Ming Li}
\email{lim@hfut.edu.cn}
\affiliation{School of Physics, Hefei University of Technology, Hefei, Anhui 230009, China}

\author{Youjin Deng}
\email{yjdeng@ustc.edu.cn}
\affiliation{Department of Modern Physics, University of Science and Technology of China, Hefei, Anhui 230026, China}
\affiliation{Hefei National Research Center for Physical Sciences at the Microscale, University of Science and Technology of China, Hefei, Anhui 230026, China}
\affiliation{Hefei National Laboratory, University of Science and Technology of China, Hefei, Anhui 230088, China}

\date{\today}

\begin{abstract}
Spatial self-similarity is a hallmark of critical phenomena. We study the dynamic process of percolation, in which bonds are incrementally added to an initially empty lattice until the system becomes fully occupied. By tracking the gap -- the size increment of clusters upon bond addition -- and the corresponding merged cluster, we identify scale-invariant temporal patterns in both quantities throughout a large portion of the process. This reveals a form of temporal self-similarity that has not been reported before. We further establish quantitative relations between the dynamic scaling exponents and the conventional static critical exponents, which enable the determination of critical behavior without prior knowledge of the critical point. The same self-similar dynamics is observed in both bond and site percolation on lattices and networks, and extends to other systems such as explosive and rigidity percolation. Moreover, similar temporal scaling is found in the initial nonequilibrium evolution of the Bak-Tang-Wiesenfeld sandpile model, suggesting a dynamic critical behavior distinct from its equilibrium state. These results provide a unified framework for understanding critical dynamics and may find applications in a broad range of complex systems.
\end{abstract}

\maketitle

\section{Introduction}

Critical phenomena describe the behaviors of systems undergoing continuous phase transitions~\cite{Ma2018}, where macroscopic properties change dramatically in response to small variations of control parameters. A key feature is spatial self-similarity, where structures at different scales appear statistically identical. Percolation is a paradigmatic model for studying critical phenomena~\cite{Stauffer1991}. As the occupation probability $p$ approaches the percolation threshold $p_c$, the correlation length diverges as $\xi \sim |p-p_c|^{-\nu}$, and self-similar spatial structures emerge up to a characteristic size $s_\xi \sim \xi^{\,d_f} \sim |p-p_c|^{-1/\sigma}$, with $d_f$ the fractal dimension. The cluster number density obeys
\begin{equation}
n(s) = s^{-\tau} \, \tilde{n}(s/s_\xi) \; ,   \label{eq-nsl}
\end{equation}
with $\tau$ the Fisher exponent and $\tilde{n}(\cdot)$ a universal scaling function. Most critical behaviors can be captured by two independent exponents, such as $\nu$ and $d_f$, from which others can be derived via scaling relations. In $d$ dimensions, $1/\sigma = \nu d_f$ and $\tau = 1 + d/d_f$. In two dimensions (2D), $\nu = 4/3$ and $d_f = 91/48$ give $\sigma = 36/91$ and $\tau = 187/91$~\cite{Nienhuis1980,Nienhuis1984,Nienhuis1987,Cardy1987,Smirnov2001}. Changing spatial dimension or cluster formation rules, such as in explosive percolation (EP)~\cite{Achlioptas2009,Riordan2011} and rigidity percolation~\cite{Jacobs1995,Jacobs1996,Moukarzel1999,Briere2007}, can alter critical exponents, but the scaling relations hold across models.

Numerical studies of critical phenomena are generally performed on finite systems, where self-similar behavior is characterized by finite-size scaling (FSS) theory. The central assumption of FSS is that, for a system of linear size $L$, when the control parameter $p$ lies within a finite-size critical window $|p-p_c|\in\mathcal{O}(L^{-1/\nu})$, the correlation length saturates at $\mathcal{O}(L)$. As a consequence, singular critical observables manifest as power-law scaling with respect to the system size $L$. For example, the size of the largest cluster $C_1$, which represents the characteristic cluster size $s_\xi$, follows the scaling relation $C_1\sim L^{d_f}$. Such FSS relations provide a standard numerical route to determine critical exponents, provided that critical observables can be accurately measured within the finite-size critical window for a series of system sizes $L$.

However, this critical window $\mathcal{O}(L^{-1/\nu})$ shrinks rapidly as $L$ increases. As a result, even small uncertainties in the estimated critical point $p_c$ can in principle shift large-scale simulations outside the true scaling regime, thereby invalidating the application of FSS. This makes FSS analyses sensitive to the precise, a priori determination of $p_c$. Moreover, even when $p_c$ is determined with high accuracy, strong finite-size corrections or large sample-to-sample fluctuations often obscure the asymptotic scaling behavior in many complex systems. For instance, EP was initially misidentified as a discontinuous phase transition due to its anomalous FSS behavior~\cite{Friedman2009,Ziff2009,Achlioptas2009,Souza2010,Radicchi2010,Grassberger2011,Souza2015}. Similarly, in 2D rigidity percolation, reliable estimates of critical exponents remain challenging because of pronounced finite-size effects~\cite{Feng1984,Kantor1984,Hansen1989,Jacobs1995,Jacobs1996,Moukarzel1999,Plischke2007,Briere2007,Zhang2015,Ellenbroek2015,Lu2024}.

These limitations reflect a more fundamental issue: conventional FSS relies on finely tuning the control parameter to a critical point and extracting scaling behavior from essentially static snapshots. In contrast, when a system evolves dynamically and the control parameter is continuously varied, it naturally traverses the critical regime and encodes rich temporal information, which is largely discarded in static analyses. Motivated by this perspective, exploring the inherent dynamic structure of cluster formation provides a complementary route to characterize critical behavior, without requiring a priori knowledge of the critical point. Such a dynamic framework also better reflects many real-world processes, ranging from power grids~\cite{Buldyrev2010} and climate dynamics~\cite{Fan2018} to traffic congestion~\cite{Li2014} and information spreading~\cite{Xie2021}.

\begin{figure*}
\centering
\includegraphics[width=2.0\columnwidth]{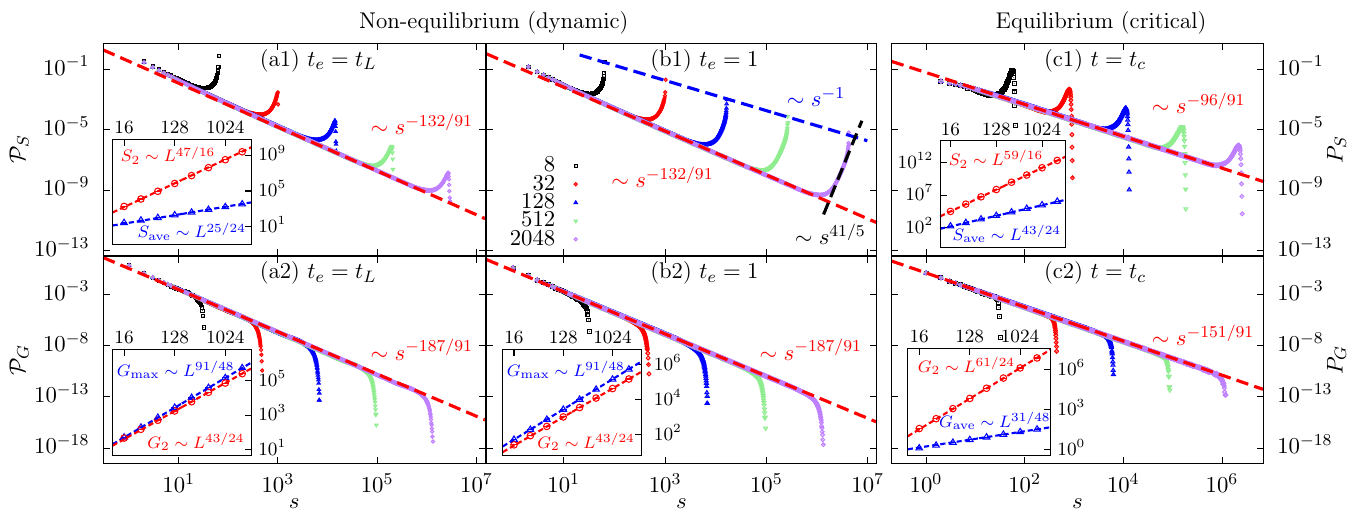}
\caption{(Color online) Temporal and spatial self-similarity in bond percolation on square lattices with side length $L = 8$-$2048$. (a) Dynamic size distributions of clusters, $\mathcal{P}_S(s,L) \sim s^{-\tau_S'}$, and of gaps, $\mathcal{P}_G(s,L) \sim s^{-\tau_G'}$, measured during the bond-insertion dynamics up to the pseudocritical point $t_L$. Insets: FSS of $S_{\rm ave}$, $S_2$, $G_2$, and $G_{\max}$. (b) Dynamic distributions measured up to $t_e = 1$. The bump in (b1): before reaching its maximum, the distribution follows a scaling of $\sim s^{\,41/5}$, while the peak height decays as $\sim s^{-1}$, corresponds to the contribution from the giant cluster. (c) Static distributions $P_S(s,L) \sim s^{-\tau_S}$ and $P_G(s,L) \sim s^{-\tau_G}$ at $t_c$, with insets verifying the FSS of $S_{\rm ave}$, $S_2$, $G_{\rm ave}$, and $G_2$. The values of the corresponding exponents are labeled in the figure, and their detailed derivations are given in Sec.~\ref{sec-ssipd}.}
\label{f1}
\end{figure*}

In this paper, we uncover a temporal self-similarity that emerges from the dynamics of cluster merging in percolation. For bond percolation, we define evolutionary time $t$ as the fraction of occupied bonds $B/E$, where $B$ is the number of inserted bonds and $E$ is the total number of bonds. This simple bond-insertion process, while monotonically related to the overall bond occupied probability $p$, reveals underlying kinetic correlations. At each step, we record the minimal size increment in clusters (termed gap) and the size of the resulting merged cluster. The temporal evolution reveals self-similar dynamics: both the gap-size distribution $\mathcal{P}_G(s,L)$ and the cluster-size distribution $\mathcal{P}_S(s,L)$ conform to the scaling form of Eq.~(\ref{eq-nsl}), see Figs.~\ref{f1} (a) and (b). The cutoff size is $s_\xi \sim L^{d_f}$, while the Fisher exponents are $\tau_G'=\tau$ for gaps and $\tau_S' = \tau + \sigma - 1$ for clusters. Dynamic observables exhibit clean FSS behaviors, e.g., the largest gap $G_{\max} \sim L^{d_f}$.

This dynamic provides a novel, universally applicable, and highly efficient framework for extracting critical properties, crucially bypassing the need for precise prior determination of the critical point $p_c$. This addresses a major technical bottleneck in conventional FSS analysis. A simple example is 1D percolation, where $p_c = 1$ precludes any critical behavior in conventional FSS analyses. However, its temporal dynamics clearly exhibits self-similarity, with both $\mathcal{P}_G(s,L)$ and $\mathcal{P}_S(s,L)$ following a power law characterized by $\tau_G' = \tau_S' = 2$, yielding $\nu = d_f = 1$. Applied in EP, this approach shows that the system still obeys the standard FSS theory, and allows for the extraction of high-precision critical exponents. Furthermore, the dynamic analysis in rigidity percolation immediately reveals a previously unreported self-similar cascade of cluster mergings~\cite{Lu2024,Lu2026}, underscoring the transformative power of the temporal self-similarity framework.

To demonstrate the broad scope and universality of this dynamic scaling, we apply the temporal framework to the Bak-Tang-Wiesenfeld (BTW) sandpile model~\cite{Bak1987,Tang1988,Engsig2025,Manna2025}, a paradigmatic example of self-organized criticality (SOC). In this model, grains are added one by one to randomly selected sites of an initially empty square lattice with open boundaries. Each site can hold at most three grains, and when a fourth grain arrives, it topples, transferring one grain to each of its four neighbors. This toppling may trigger further topplings, leading to a cascade of events called an avalanche.

We focus on the early, transient dynamics leading up to the first spanning avalanche, before the system settles into the stationary SOC state. At each step, we record both the avalanche size $S$ (the total number of topplings) and the avalanche area $A$ (the number of distinct sites toppled). Both the two distributions ${\cal P}_S$ and ${\cal P}_A$ display clear power-law forms. The discovery that this same functional form of temporal self-similarity exists in the non-equilibrium BTW model strongly supports the hypothesis of a broad, underlying temporal universality in critical dynamics, regardless of whether the system is externally tuned or self-organized. Interestingly, while the avalanche area $A$ is governed by a single fractal dimension, the avalanche size $S$ does not collapse with a single exponent, suggesting a richer, multifractal behavior akin to that reported in earlier studies~\cite{Tebaldi1999}. This temporal approach thus provides a fresh lens for understanding the complex scaling properties of SOC dynamics.

The remainder of this paper is organized as follows. In Sec.~\ref{sec-gdo}, we introduce the gap-dynamics framework and define the key observables.
Sections~\ref{sec-ssipd} and \ref{sec-bps} present the theoretical predictions and numerical results for various percolation systems. Section~\ref{sec-sand} demonstrates the applicability of the dynamic framework to the BTW sandpile model. Finally, Sec.~\ref{sec-dis} concludes with a brief discussion.

\section{Gap dynamics and observables}  \label{sec-gdo}

\subsection{Dynamic bond-addition process}

We introduce a unified dynamic framework applicable to a broad class of percolation-type systems, including bond, site, explosive, and rigidity percolation on lattices and networks. The system starts from an empty configuration and evolves through the sequential addition of elementary elements, such as bonds or sites, one at a time. After each addition, the newly introduced element may connect multiple existing clusters, resulting in the formation of a larger merged cluster. The number of added elements (bonds or sites) is denoted as $B$. This process continues until a tunable endpoint $B_e$, corresponding to full occupation when $B_e$ equals the total number of bonds $E$ or sites $V$ in the system.

To facilitate comparison across different systems, we define a normalized time-like variable
\begin{equation}
t \equiv \frac{B}{E}, \quad \text{or} \quad t \equiv \frac{B}{V},
\end{equation}
depending on whether bonds or sites are added. The corresponding endpoint is denoted by $t_e$, with a maximum value of $1$ corresponding to the fully occupied state. Numerically, the dimensionless variable $t \in [0,1]$ is equivalent to the bond/site occupation probability $p$. However, they play distinct conceptual roles: while $p$ characterizes a static configuration at a given occupation level, $t$ parameterizes the dynamical evolution of the system through successive addition events and thus serves as a natural time-like variable in the dynamic process.

For Bernoulli bond or site percolation, an analogous dynamic process underlies the celebrated Newman-Ziff algorithm~\cite{Newman2001a}. Moreover, similar dynamic constructions naturally arise in a broad class of percolation-related systems, including EP~\cite{Achlioptas2009} and growing graphs~\cite{Dorogovtsev2001,Bollobas2004}.

\subsection{Observables}

At each merging event during the dynamic process, we characterize the structural change using two quantities: the \emph{merged cluster size} $\mathcal{S}$ and the \emph{gap size} $\mathcal{G}$. The merged cluster size
\begin{equation}
\mathcal{S} = \sum_i s_i
\end{equation}
measures the size of the cluster formed after a merging event, where ${s_i}$ denote the sizes of the clusters involved. To quantify the effective impact of the merging operation itself, we define the gap size as
\begin{equation}
\mathcal{G} = \mathcal{S} - s_{\rm max},   \label{eq-dg}
\end{equation}
which corresponds to the contribution from all but the largest participating cluster $s_{\rm max}$. By excluding the trivial contribution from a dominant cluster, $\mathcal{G}$ isolates the genuine structural change induced by the merging event and remains sensitive to the dynamics of finite clusters even in the presence of a giant cluster. If the added element is internal, i.e., it does not merge distinct clusters, then $\mathcal{G}=0$ and $\mathcal{S}$ reduces to the size of the cluster involved.

During the dynamic process, the gap size $\mathcal{G}(t)$ typically exhibits an overall growth-decay trend: it increases at early stages, reaches a maximum value $\mathcal{G}_{\rm max}$ at a characteristic dynamic pseudocritical time $t_L$, and then decreases, while displaying fluctuations at the level of individual merging events. The ensemble-averaged quantities $G_{\rm max}\equiv\langle \mathcal{G}_{\rm max}\rangle$, together with the mean and variance of $t_L$, naturally characterize the dynamic critical behavior and form the basis of the FSS analysis~\cite{Li2023,Li2024,Li2024a,Fan2020}.

In this work, we are interested in the probability distribution of gap sizes during the dynamic process. We sample the distribution $\mathcal{P}_G(s,L)$, where $\mathcal{P}_G(s,L)\Delta s$ denotes the normalized count of gap events with sizes in the interval $(s-\Delta s/2,s+\Delta s/2]$. To properly capture the broad range of gap sizes, the bin width $\Delta s$ is chosen to increase as a power law of $s$. While $\mathcal{P}_G(s,L)$ can in principle be obtained from a single stochastic realization, reliable statistics generally require averaging over many independent realizations.

The average and squared-average of the gap size over an entire dynamic process are then given by the first and second moments of $\mathcal{P}_G(s,L)$,
\begin{align}
G_{\rm ave}(L) &= \sum_{s} s\mathcal{P}_G(s,L) = \left\langle \frac{1}{B_e} \sum_{B=1}^{B_e} \mathcal{G}(B) \right\rangle, \\
G_2(L) &= \sum_{s} s^2\mathcal{P}_G(s,L) = \left\langle \frac{1}{B_e} \sum_{B=1}^{B_e} \mathcal{G}^2(B) \right\rangle,
\end{align}
where $\langle\cdot\rangle$ denotes an average over independent realizations ($10^5$–$10^6$ for typical system sizes). Analogous quantities $S_{\rm ave}$, $S_2$, and the dynamic distribution $\mathcal{P}_S(s,L)$ are defined for the merged cluster size $\mathcal{S}$.

It is natural to probe how a critical configuration responds to a local bond or site addition, which provides a susceptibility-like measure of the instantaneous response of critical states to a minimal structural perturbation. At the percolation threshold $t_c=p_c$, we therefore consider the effect of adding a single bond or site at random to a critical configuration and measure the resulting quantities $\mathcal{G}$ and $\mathcal{S}$. This single-addition operation acts as a local probe of the critical structure, and the added bond or site is not retained after the measurement. Repeating this procedure over many independent critical configurations yields a large ensemble of values of $\mathcal{G}$ and $\mathcal{S}$, which characterize the instantaneous response of critical states. For notational convenience, we refer to the corresponding distributions as the \emph{static} distributions and denote them by $P_G(s,L)$ and $P_S(s,L)$, using normal fonts to distinguish them from the dynamic distributions $\mathcal{P}_G(s,L)$ and $\mathcal{P}_S(s,L)$ obtained along the evolving process.

We also note that the cluster-number density $n(s,L)$, a standard static quantity, is defined as the number of clusters of size $s$ per unit volume $V=L^d$ in a configuration, and should be distinguished from the static distribution $P_S(s,L)$.

\section{Self-similarity in percolation dynamics}   \label{sec-ssipd}

\subsection{Static critical configuration}

Similar to the cluster-number density $n(s,L)$ in Eq.~(\ref{eq-nsl}), the static distributions at criticality should obey the FSS forms
\begin{equation}
P_X(s, L) = s^{-\tau_X} \, \tilde{P}_X(s/L^{d_f})
\end{equation}
where $X$ denotes either the gap size $G$ or the cluster size $S$. Thus, the first moment of $P_X(s, L)$ gives the scaling forms of the average sizes of cluster and gap,
\begin{align}
S_{\rm ave} &\sim L^{(2-\tau_S)\,d_f}, \label{eq-sa}  \\
G_{\rm ave} &\sim L^{(2-\tau_G)\,d_f}. \label{eq-ga}
\end{align}

By definition, for a static configuration the size $\mathcal{S}$ equivalently represents the cluster size accessed by a randomly chosen bond or site. Its ensemble average $S_{\rm ave}$ corresponds to the magnetic susceptibility of the percolation transition, i.e., the second moment of the cluster-number density $n(s,L)$~\cite{Stauffer1991}, with the known FSS
\begin{equation}
S_{\rm ave} \sim L^{2 d_f - d}, \label{eq-sa0}
\end{equation}
where $d$ is the system dimension. Similarly, the average gap size $G_{\rm ave}$ corresponds to the typical size of the smaller cluster connected by a randomly chosen bond. Equivalently, for site percolation, it measures all clusters accessed by a site except the largest one. Its FSS form has been derived in Ref.~\cite{Deng2010},
\begin{equation}
G_{\rm ave} \sim L^{d_f-d+1/\nu}, \label{eq-ga0}
\end{equation}
where $\nu$ is the correlation-length exponent. Comparing Eqs.~(\ref{eq-sa}) and (\ref{eq-ga}) with Eqs.~(\ref{eq-sa0}) and (\ref{eq-ga0}) leads to the relations between the Fisher exponents
\begin{align}
\tau_S &= \tau - 1,   \label{eq-taus} \\
\tau_G &= \tau - \sigma, \label{eq-taug}
\end{align}
where $\tau = 1 + d/d_f$ is the standard Fisher exponent as in Eq.~(\ref{eq-nsl}) and $\sigma = 1/\nu d_f$ is the critical exponent of the characteristic cluster size $s_\xi$. Since the probability that a randomly chosen bond or site belongs to a cluster of size $s$ is proportional to $s$, the relation $P_S(s,L)\sim s\,n(s,L)$ holds, recovering Eq.~(\ref{eq-taus}). These relations show how the static gap and cluster distributions encode the standard Fisher exponents of percolation, connecting the single-bond or -site response to the global critical scaling.

In 2D, with $d_f = 91/48$ and $\nu=4/3$, one has $\tau = 187/91$ and $\sigma = 36/91$, yielding $\tau_S = 96/91$ and $\tau_G = 151/91$. Figures~\ref{f1} (c1) and (c2) show the simulation results for the static cluster-size and gap-size distributions $P_G(s,L)$ and $P_S(s,L)$ obtained from bond percolation on square lattices. The measured distributions follow the predicted power-law behaviors $P_G\sim s^{-\tau_G}$ and $P_S\sim s^{-\tau_S}$ over a broad range, in excellent agreement with the theoretical exponents. The insets present the FSS of their moments, confirming the expected scaling: $S_{\rm ave} \sim L^{(2-\tau_S)d_f} \sim L^{43/24}$, $G_{\rm ave} \sim L^{(2-\tau_G)d_f} \sim L^{31/48}$, $S_2 \sim L^{(3-\tau_S)d_f} \sim L^{59/16}$, and $G_2 \sim L^{(3-\tau_G)d_f} \sim L^{61/24}$. These results provide a clear verification of the static critical scaling and validate the Fisher exponents derived above.

\subsection{Dynamic process}

Figures~\ref{f1} (a1,a2) and (b1,b2) show the dynamic distributions $\mathcal{P}_S(s, L)$ and $\mathcal{P}_G(s, L)$ for bond percolation on square lattices, measured up to $t_e = t_L$ and $t_e = 1$, respectively. Both distributions exhibit clear algebraic scaling, indicating dynamic self-similarity. However, the associated Fisher exponents differ from their static counterparts, with $\tau_S' = 132/91$ and $\tau_G' = 187/91$. Careful analysis shows that these dynamic exponents satisfy a simple shift relation (to be derived later)
\begin{align}
\tau_S' &= \tau_S + \sigma = \tau + \sigma - 1, \label{eq-tauds}  \\
\tau_G' &= \tau_G + \sigma = \tau. \label{eq-taudg}
\end{align}

The fact that $\tau_G' = \tau$ is highly interesting. It implies that the dynamic gap-size distribution obeys the same clean scaling as the static cluster-number density at criticality, even beyond the percolation threshold ($t_e > t_L$). As shown in the inset, its moments also follow the same FSS behavior as those of the static cluster-number density. This provides a powerful framework to extract critical properties without prior knowledge of the precise critical point. Such robustness originates from the temporal accumulation inherent in the dynamic process and is consistent with the general framework of critical phenomena. A more detailed explanation will be given later.

In contrast, the dynamic cluster-size distribution $\mathcal{P}_S(s, L)$ exhibits a Fisher exponent $\tau_S' = \tau + \sigma - 1 = (d + 1/\nu)/d_f$, which encodes information from both the fractal dimension $d_f$ and the correlation-length exponent $\nu$. This reflects the richer structure of temporal accumulation in $\mathcal{P}_S$, incorporating contributions from both the critical regime and the approach-to-critical stages of the dynamic process.

Furthermore, when the process is terminated in the supercritical phase (e.g., at $t_e = 1$), $\mathcal{P}_S(s, L)$ acquires an additional contribution from the giant cluster $C_1 \sim L^d$, reflecting the nonzero probability that a bond belongs to the largest cluster in the supercritical regime. In Fig.~\ref{f1}(b1), this contribution manifests as a distinct bump for large $s$, with its height decaying as $\sim s^{-1}$. This feature can also be interpreted within the standard framework of phase transition theory and will be derived in detail later. In contrast, the gap-size distribution $\mathcal{P}_G(s, L)$ remains free from such a bump, since the definition of the gap naturally excludes any contribution from the giant cluster, see Fig.~\ref{f1}(b2).

\subsection{Derivation of dynamic Fisher exponents}   \label{sec-ddf}

We now derive the dynamic Fisher exponent relations, Eqs.~(\ref{eq-tauds}) and (\ref{eq-taudg}), within the framework of phase transition and FSS theories. The key idea is that the dynamic distributions can be regarded as the temporal accumulation of the corresponding static ones. Given an evolution time $t$ for the dynamic process, as long as the instantaneous correlation length $\xi(t)$ is much larger than the lattice spacing, each static observable at time $t$ follows the critical scaling form characterized by relations such as $\xi(t) \sim |t - t_c|^{-\nu}$ and $P_X(s,\xi(t)) \sim s^{-\tau_X} \tilde{P}_X(s/s_\xi(t))$. Their accumulation over time thus naturally preserves the scaling behavior, leading to well-defined dynamic exponents.

\subsubsection{Universal dynamic scaling of finite clusters}

We begin by considering the process that terminates near the critical point ($t_e \approx t_c$). In the thermodynamic limit ($L \to \infty$), the bond- (or site-) addition dynamics can be viewed as a continuous realization of the growth of the correlation length $\xi(t)$. The dynamic distribution $\mathcal{P}_X(s)$ can therefore be regarded as a weighted superposition of the static distributions $P_X(s,\xi)$ over all possible correlation lengths $\xi(t)$ sampled during the evolution. Since the dynamic process proceeds uniformly in time, the weighting factor $W(\xi)$ for a static distribution $P_X(s,\xi)$ can be directly characterized by the corresponding time interval $dt$, leading to
\begin{equation}
W(\xi)\, d\xi = dt.   \label{eq-wxi}
\end{equation}

At each time $t$ with $\xi(t)\gg 1$, the system behaves as if it were in a quasi-critical state characterized by the correlation length $\xi \sim |t-t_c|^{-\nu}$. Differentiating $\xi \sim |t-t_c|^{-\nu}$ yields
\begin{align}
d\xi &\sim |t - t_c|^{-\nu-1} dt   \\
     &\sim \xi^{\,1 + 1/\nu} dt.
\end{align}
Substituting to Eq.~(\ref{eq-wxi}), we have
\begin{equation}
W(\xi) \sim \xi^{\, -1 - 1/\nu }.
\end{equation}
Recasting this expression in terms of the cutoff size $s_\xi \sim \xi^{\,d_f}$ gives
\begin{equation}
W(s_\xi) \sim s_\xi^{\,-\sigma - 1}.
\end{equation}
This shows that the weighting factor $W(s_\xi)$ itself obeys scaling with respect to the cutoff size $s_\xi$ or the correlation length $\xi$.

The dynamic distribution $\mathcal{P}_X(s)$ is therefore obtained as the integral of the static distribution $P_X(s,s_\xi)$ weighted by $W(s_\xi)$, over all relevant cutoff sizes $s_\xi \ge s$,
\begin{align}
\mathcal{P}_X(s) & \sim \int_s^{\infty} W(s_\xi) P_X(s,s_\xi)\, ds_\xi \\
& \sim s^{-\tau_X} \int_s^{\infty} s_\xi^{-\sigma-1} \tilde{P}_X(s/s_\xi)\, ds_\xi.  \label{eq-pxi}
\end{align}
This integral is simplified by the asymptotic behavior of the scaling function: $\tilde{P}_X(s/s_\xi)$ approaches a constant near $s_\xi = s$ and vanishes rapidly as $s_\xi \to \infty$. Consequently, we have
\begin{equation}
\mathcal{P}_X(s) \sim s^{-\tau_X - \sigma}.  \label{eq-px}
\end{equation}
Therefore, the dynamic Fisher exponents satisfy
\begin{equation}
\tau_X' = \tau_X + \sigma,
\end{equation}
confirming Eqs.~(\ref{eq-tauds}) and (\ref{eq-taudg}).

\subsubsection{Dynamic contribution of the giant cluster}

When the dynamic process proceeds into the supercritical regime ($t_e > t_c$), a macroscopic (giant) cluster emerges and continues to grow. As a result, the overall dynamic cluster-size distribution $\mathcal{P}_S(s)$ inevitably acquires contributions from this giant cluster when the dynamic statistics are accumulated over the broader critical region ($\xi \gg 1$). In contrast, by construction, the gap distribution $\mathcal{P}_G(s)$ naturally excludes the giant cluster and therefore retains the scaling form derived above.

In the thermodynamic limit ($L \to \infty$), the number density of the giant cluster in a static configuration at $t>t_c$ can be modeled by a Dirac delta function
\begin{equation}
n_1(s) =\frac{1}{V} \delta(s - C_1),
\end{equation}
where $C_1$ is the size of the largest (giant) cluster, and $V=L^d$ is the system volume. The normalization $V\int n_1(s)\, ds = 1$ ensures the uniqueness of the giant cluster. For a randomly inserted bond or site, the probability that it belongs to the giant cluster is $m\equiv C_1/V$, yielding the static size distribution of the giant cluster
\begin{align}
P_1(s) &= m \, s \, n_1(s)  \nonumber \\
 &= \frac{s C_1}{V^2}\, \delta(s - C_1).
\end{align}
Here, the additional factor $s$ accounts for the fact that, when a bond or site is chosen at random, the probability of sampling a cluster of size $s$ is proportional to its size -- larger clusters contain more sites or bonds and are thus more likely to be selected. This is consistent with the general relation $P_S(s) = s\,n(s)$ introduced earlier.

The dynamic process beyond $t_c$ can be viewed as a temporal accumulation of configurations with an increasing giant cluster size $C_1(t)$. The corresponding dynamic distribution $\mathcal{P}_1(s)$ can then be obtained by integrating $P_1(s)$ over $C_1 \in [1, V\to \infty)$ with an $C_1$-dependent weight $W(C_1)$. Similar to Eq.~(\ref{eq-wxi}), $W(C_1)$ is also characterized by the time interval $dt$,
\begin{equation}
W(C_1)\, dC_1 \sim dt.
\end{equation}
Note that the probability $m(t)\equiv C_1(t)/V$ is just the order parameter of percolation transition. When $\xi \gg 1$, it obeys the scaling $m(t) \sim (t - t_c)^{\,\beta}$ for $t>t_c$, so that $C_1(t)\sim V (t - t_c)^{\,\beta}$. This yields
\begin{align}
dC_1 &\sim V (t - t_c)^{\,\beta-1} dt \\
&\sim V^{1/\beta} C_1^{1 - 1/\beta} dt,
\end{align}
leading to
\begin{equation}
W(C_1) \sim V^{\,-1/\beta} C_1^{1/\beta-1} .
\end{equation}
Therefore, the dynamic distribution $\mathcal{P}_1(s)$ of the giant cluster can be expressed as
\begin{align}
\mathcal{P}_1(s) & \sim \int_1^{\infty} W(C_1) P_1(s)\, dC_1  \nonumber \\
& \sim V^{-2 - 1/\beta} s \int_1^{\infty} C_1^{1/\beta} \delta(s - C_1)\, dC_1   \nonumber \\
& \sim V^{-2 - 1/\beta} s^{1 + 1/\beta}.   \label{eq-sv}
\end{align}
This expression applies only to the giant-cluster sector $s>s_\xi$, where $s$ is of order $C_1$.

Consequently, the total dynamic cluster-size distribution under $t_e > t_c$ can be expressed as a composite form
\begin{equation}
\mathcal{P}_S(s) = s^{-(\tau + \sigma - 1)} \tilde{\mathcal{P}}_S(s/s_\xi) + V^{-2 - 1/\beta} s^{1 + 1/\beta} \tilde{\mathcal{P}}_1(s/s_\xi),   \label{eq-psall}
\end{equation}
where the scaling function $\tilde{\mathcal{P}}_S(s/s_\xi)$ is nearly constant for $s < s_\xi$ and vanishes rapidly otherwise, while $\tilde{\mathcal{P}}_1(s/s_\xi)$ behaves oppositely. Physically, the two contributions in Eq.~(\ref{eq-psall}) are statistically dominated by different stages of the dynamic process. The first term is dominated by configurations near and below $t_c$, since for $t>t_c$ a randomly added bond or site is increasingly likely to involve the giant cluster, making events that connect only finite clusters rare. In contrast, the second term arises entirely from the supercritical regime and reflects the temporal accumulation of the growing giant cluster.

When $s > s_\xi$, the first term in Eq.~(\ref{eq-psall}) vanishes. As a result, for $t_e > t_c$, the tail of the dynamic cluster-size distribution is governed by the contribution of the giant cluster and behaves as Eq.~(\ref{eq-sv}). For a finite system with fixed volume $V$, this implies an algebraic increase $\mathcal{P}_S(s) \sim s^{1 + 1/\beta}$ in the large-$s$ regime. Consequently, $\mathcal{P}_S(s)$ grows with $s$, giving rise to the hump observed at large $s$ in Fig.~\ref{f1} (b1). For 2D percolation, $\beta = 5/36$, yielding $\mathcal{P}_S(s) \sim s^{41/5}$, in agreement with the numerical results in Fig.~\ref{f1} (b1). This growth persists until $s$ approaches the system-size cutoff $s \sim V$, corresponding to the largest possible cluster in the dynamic process. Substituting $s \sim V$ into Eq.~(\ref{eq-sv}) yields $\mathcal{P}_S(s) \sim V^{-1} \sim s^{-1}$, indicating that the height of the hump decays as $\sim s^{-1}$. Notably, this decay is independent of the critical exponents and thus universal, as confirmed by Fig.~\ref{f1} (b1) and additional examples presented later.

\section{Self-similar dynamics in broad percolation systems}    \label{sec-bps}

Beyond providing an alternative route to reproduce known critical properties, the dynamic approach offers a general framework to characterize criticality from a temporal perspective. In particular, it is applicable to systems where the static ensemble description is problematic or where the critical point $p_c$ cannot be precisely determined. In the following, we present representative examples that illustrate the generality and robustness of this dynamic approach.

\begin{figure}
\centering
\includegraphics[width=1.0\columnwidth]{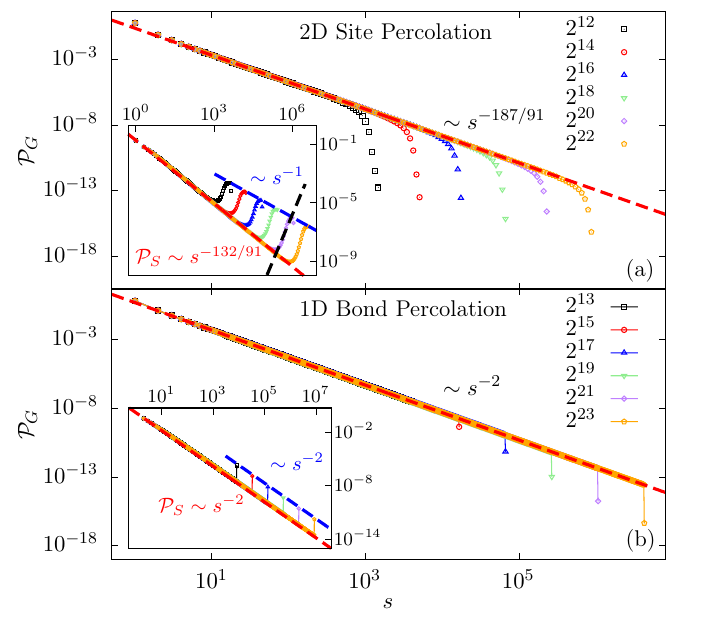}
\caption{(Color online) Dynamic gap-size distribution $\mathcal{P}_G(s,L)$ in 2D and 1D percolation with $t_e=1$. Insets show the corresponding dynamic cluster-size distribution $\mathcal{P}_S(s,L)$. (a) Site percolation on square lattices for different $V=L^2$. (b) Bond percolation in 1D for different $V=L$. In the two cases, robust power-law scalings $\mathcal{P}_G \sim s^{-\tau_G'}$ and $\mathcal{P}_S \sim s^{-\tau_S'}$ are observed (red dashed line), with $\tau_G'$ and $\tau_S'$ consistent with Eqs.~(\ref{eq-tauds}) and (\ref{eq-taudg}). For 2D percolation, $\mathcal{P}_S(s,L)$ exhibits the predicted scaling $\sim s^{1+1/\beta}\sim s^{\,41/5}$ for large $s$ (black dashed line), followed by a vanishing hump height $\sim s^{-1}$ (blue dashed line), in agreement with Eq.~(\ref{eq-sv}). For 1D percolation, the supercritical regime is absent, so $\mathcal{P}_S(s,L)$ behaves effectively as if $t_e = t_c$, with the bump height scaling as $\sim s^{-\tau_S'}$.}
\label{f2}
\end{figure}

\subsection{Site percolation in 2D}

For site percolation, the dynamic process proceeds by occupying sites randomly one by one. Unlike bond percolation, adding a single site can merge multiple clusters simultaneously. Following Eq.~(\ref{eq-dg}), the gap is defined as the total size of all merged clusters excluding the largest one.

Figure~\ref{f2}~(a) shows simulation results on square lattices. Both the dynamic gap-size distribution, $\mathcal{P}_G(s,L) \sim s^{-\tau_G'}$, and the dynamic cluster-size distribution, $\mathcal{P}_S(s,L) \sim s^{-\tau_S'}$ (inset), exhibit clear power-law scaling with exponents $\tau_G' = 187/91$ and $\tau_S' = 132/91$. These values coincide with those obtained for bond percolation in 2D (Fig.~\ref{f1}), confirming the universality of the dynamic scaling across site and bond percolation. The inset further shows that, for $t_e=1$, a bump appears for large $s$, contributed by the giant cluster. As predicted by Eq.~(\ref{eq-sv}), a scaling $\sim s^{1 + 1/\beta} \sim s^{41/5}$ exists for large $s$, with the bump height scaling as $\sim s^{-1}$.

\subsection{Bond percolation in 1D}

In 1D percolation, the threshold is trivially located at $t_c = 1$, where all sites become connected, so no conventional critical behavior is observed. Nevertheless, as shown in Fig.~\ref{f2}(b), the dynamically obtained gap-size distribution during the bond-insertion process still follows a clear power law, $\mathcal{P}_G(s,L) \sim s^{-\tau_G'}$, with $\tau_G' = 2$, consistent with the known Fisher exponent $\tau = 1 + d/d_f = 2$. This reflects the continuous nature of the transition with $1/\nu = d_f = 1$.

Since $\sigma = 1$ in 1D, the corresponding dynamic cluster-size distribution $\mathcal{P}_S(s,L)$ exhibits the same Fisher exponent, $\tau_S' = \tau + \sigma - 1 = 2$, as shown in the inset of Fig.~\ref{f2}(b). Because $t_c = 1$, the system remains effectively subcritical even as $t_e = 1$, and thus the bump in $\mathcal{P}_S(s,L)$ at large $s$ scales as $\sim s^{-\tau_S'} \sim s^{-2}$, in contrast to the scaling $\sim s^{-1}$ observed in models with $t_c<1$.

\begin{figure}
\centering
\includegraphics[width=1.0\columnwidth]{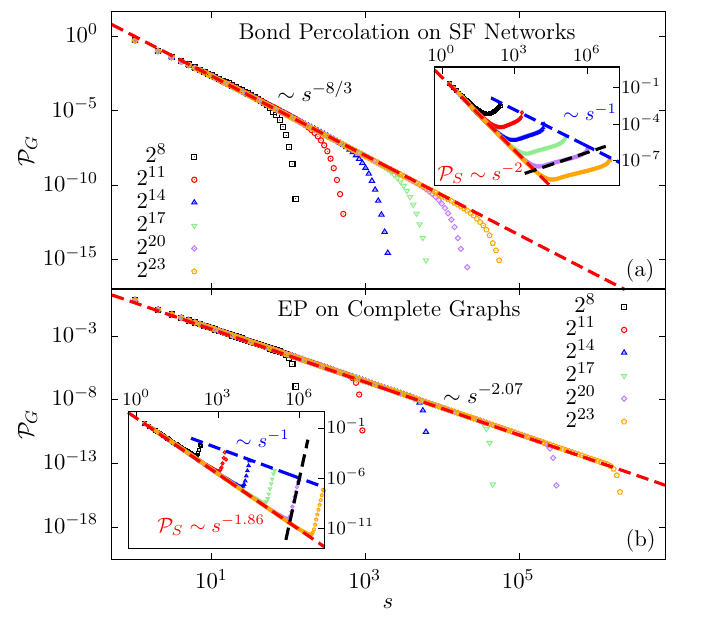}
\caption{(Color online) Dynamic gap-size distribution $\mathcal{P}_G(s,L)$ in percolation on networks with $t_e=1$. Insets show the corresponding dynamic cluster-size distribution $\mathcal{P}_S(s,L)$. (a) Bond percolation on SF networks with $\lambda=3.5$ for different $V$. (b) EP on complete graphs for different $V$. In the two cases, robust power-law scalings $\mathcal{P}_G \sim s^{-\tau_G'}$ and $\mathcal{P}_S \sim s^{-\tau_S'}$ are observed (red dashed line), with $\tau_G'$ and $\tau_S'$ consistent with Eqs.~(\ref{eq-tauds}) and (\ref{eq-taudg}). For large $s$, $\mathcal{P}_S(s,L)$ exhibits the predicted scaling $\sim s^{1+1/\beta}$ (black dashed line, $\sim s^{1.5}$ for (a), and $\sim s^{12.38}$ for (b)), followed by a vanishing hump height $\sim s^{-1}$ (blue dashed line), in agreement with Eq.~(\ref{eq-sv}).}
\label{f3}
\end{figure}

\subsection{Bond percolation on scale-free networks}

A scale-free (SF) network is characterized by a degree distribution $p_k \sim k^{-\lambda}$, where $k$ denotes the degree of a randomly chosen node, i.e., the number of links connected to the node. For small values of $\lambda$, the network is highly heterogeneous and contains hubs with very large degrees, whereas for large $\lambda$ the degree distribution decays rapidly, and high-degree nodes become increasingly rare. Percolation on SF networks exhibits critical behavior that depends sensitively on the degree exponent $\lambda$~\cite{Cohen2002,Lee2004,Cirigliano2024}. However, it is difficult to achieve precise numerical confirmation, due to strong finite-size corrections~\cite{Wu2007,Zhao2025}. In simulations, SF networks can be generated using the configuration model~\cite{Molloy1995,Newman2001}, in which each node is first assigned a degree drawn randomly from degree distribution $p_k$, and links are then randomly connected under the constraints of no self-loops or multiple edges.

As an example, we focus on $\lambda > 3$, where a finite-threshold continuous transition exists and the critical exponents are well established. Analytically, $\tau = (2\lambda - 3)/(\lambda - 2)$ and $\sigma = (\lambda - 3)/(\lambda - 2)$~\cite{Cohen2002,Lee2004,Cirigliano2024}; for $\lambda = 3.5$, this gives $\tau = 8/3$ and $\sigma = 1/3$. As shown in Fig.~\ref{f3}(a), the dynamically measured gap-size distribution follows $\mathcal{P}_G(s,L) \sim s^{-\tau_G'}$ with $\tau_G' = 8/3$, in agreement with the theoretical Fisher exponent $\tau$.

For the dynamic cluster-size distribution $\mathcal{P}_S(s,L)\sim s^{-\tau_S'}$, substituting the exponents $\tau = (2\lambda - 3)/(\lambda - 2)$ and $\sigma = (\lambda - 3)/(\lambda - 2)$ into $\tau_S'= \tau + \sigma - 1$ yields $\tau_S' = 2$, independent of $\lambda$. The numeric result of the case $\lambda=3.5$ in the inset of Fig.~\ref{f3}(a) is consistent well with this scaling. It is pointed out that this exponent $\tau_S' = 2$ coincides with the mean-field result for percolation on complete graphs, for which the mean-field values $\tau = 5/2$ and $\sigma = 1/2$ also gives $\tau_S' = 2$. This indicates that certain mean-field characteristics emerge for $\lambda > 3$. Note that it is generally recognized that percolation on SF networks with $\lambda \geq 4$ exhibits the standard mean-field critical behavior~\cite{Cohen2002}.

Furthermore, from Eq.~(\ref{eq-sv}), the large-$s$ regime in $\mathcal{P}_S(s,L)$ follows the scaling $\sim s^{1+1/\beta}$, which reduces to $\sim s^{\lambda - 2}$ with $\beta = 1/(\lambda - 3)$. The black dashed line in the inset of Fig.~\ref{f3}(b) clearly demonstrates this scaling $\sim s^{1.5}$ for $\lambda=3.5$. The scaling $\sim s^{-1}$ for the height of humps is also confirmed in the inset (blue dashed line).

\subsection{Explosive percolation on complete graphs}

We next apply the gap-dynamics framework to EP on complete graphs~\cite{Friedman2009,Ziff2009,Achlioptas2009,Souza2010,Radicchi2010,Grassberger2011,Souza2015,Li2023,Li2024}. EP follows the Achlioptas process~\cite{Achlioptas2009}, where at each step two candidate bonds are randomly selected and the bond minimizing the product of the sizes of the clusters it connects is added. Static FSS analyses often display anomalies reminiscent of discontinuous transitions; for a review, see Ref.~\cite{D’Souza2019}. Although EP is now understood to be continuous, extracting critical behaviors accurately remains challenging. Recently, two of us (ML and YD) have shown that EP obeys standard FSS when an event-based ensemble is applied~\cite{Li2023,Li2024}.

Applying the gap-dynamics framework, both the gap-size and cluster-size distributions exhibit clear power-law behavior (Fig.~\ref{f3}(b)). Using previously reported values for the fractal dimension $d_f = 0.935$ and the correlation-length exponent $1/\nu = 0.740$~\cite{Li2023}, which are defined with respect to the system volume $V$, we obtain $\tau = 1 + 1/d_f \approx 2.07$ and $\sigma = 1/\nu d_f \approx 0.79$. As shown in Fig.~\ref{f3}(b) and its inset, the dynamically extracted exponents, $\tau_G' \approx 2.07$ and $\tau_S' \approx 1.86$, satisfy the relations $\tau_G' = \tau$ and $\tau_S' = \tau + \sigma - 1$. Moreover, from $\beta = (1-d_f)\nu$, the scaling for large $s$ is $\sim s^{1+1/\beta} \sim s^{12.38}$. As shown in the inset of Fig.~\ref{f3}(b), there is good agreement between the simulation results and both the scaling $\sim s^{12.38}$ at large $s$ (black dashed line) and the decay $\sim s^{-1}$ of the hump height (blue dashed line).

These results demonstrate the validity and robustness of the dynamic approach, even in systems exhibiting anomalous scaling under the conventional ensemble. Notably, the bond-insertion rule in EP endows it with an intrinsically dynamic character, making the dynamic ensemble a natural and informative perspective.

\subsection{Rigidity percolation in 2D}

\begin{figure}
\centering
\includegraphics[width=1.0\columnwidth]{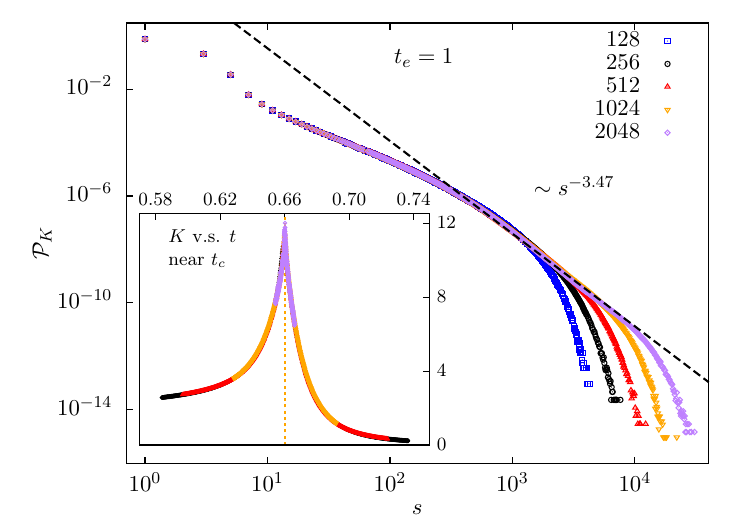}
\caption{(Color online) Distribution $\mathcal{P}_K(s,L)$ of the number of clusters merged per bond insertion during gap dynamics in rigidity percolation on triangular lattices for different $L$. As the system size $L$ increases, the distribution exhibits a clear power-law behavior, $\mathcal{P}_K(s,L) \sim s^{-\tau_K}$, with $\tau_K \approx 3.47$. The inset shows the mean number of merged clusters per step, $K(t) \equiv \langle \mathcal{K}(t) \rangle$, as a function of time $t$. Curves for different $L$ collapse onto a single, size-independent profile, with a peak at $t_c = 0.6602778(10)$ (dashed line).}
\label{f4}
\end{figure}

Rigidity percolation follows the same bond-insertion rule as ordinary bond percolation, but focuses on rigid clusters capable of transmitting mechanical stress. It has been applied to glasses, gels, amorphous solids, jamming, fibrous networks, and living tissues~\cite{Thorpe2000,Broedersz2011,Zhang2019,Rouwhorst2020,Rouwhorst2020a,Zaccone2011,Henkes2016,Dashti2023,Vinutha2023,Petridou2021,Rozman2024}. Despite its broad relevance, critical properties of rigidity percolation remain elusive, with debated critical exponents and universality classes~\cite{Feng1984,Kantor1984,Hansen1989,Jacobs1995,Jacobs1996,Moukarzel1999,Plischke2007,Briere2007,Zhang2015,Ellenbroek2015}.

In addition to exhibiting well-defined gap scaling as in the above percolation models, the gap-dynamics framework here serves mainly as an illustrative example; a detailed analysis of rigidity percolation is presented in Ref.~\cite{Lu2026}. In rigidity percolation, the gap-dynamics framework uncovers a cascade cluster-merging process: the insertion of a single bond can rigidify a nonrigid subgraph by activating multiple previously nonrigid links, triggering a collective transition. This process is quantified by $\mathcal{K}$, the number of clusters merged when a bond is inserted. As shown in Fig.~\ref{f4}, the $\mathcal{K}$ distribution $\mathcal{P}_K(s,L) \sim s^{-\tau_K}$ in an entire dynamic process follows a clear power law with $\tau_K \approx 3.47$, also revealing temporal self-similarity.

The average number of clusters merged, $K(t) \equiv \langle \mathcal{K}(t) \rangle$, exhibits a pronounced peak at the critical point $t_c$ (inset of Fig.~\ref{f4}). While the peak does not diverge and curves for different $L$ collapse without further rescaling, its location serves as a pseudocritical point. Analyzing the FSS of this pseudocritical point yields $t_c = 0.6602778(10)$, improving the precision by roughly two orders of magnitude compared to previous estimates $t_c = 0.6602(3)$~\cite{Jacobs1995,Jacobs1996}.

\section{Dynamic self-similarity in sandpile}    \label{sec-sand}

\begin{figure}
\centering
\includegraphics[width=1.0\columnwidth]{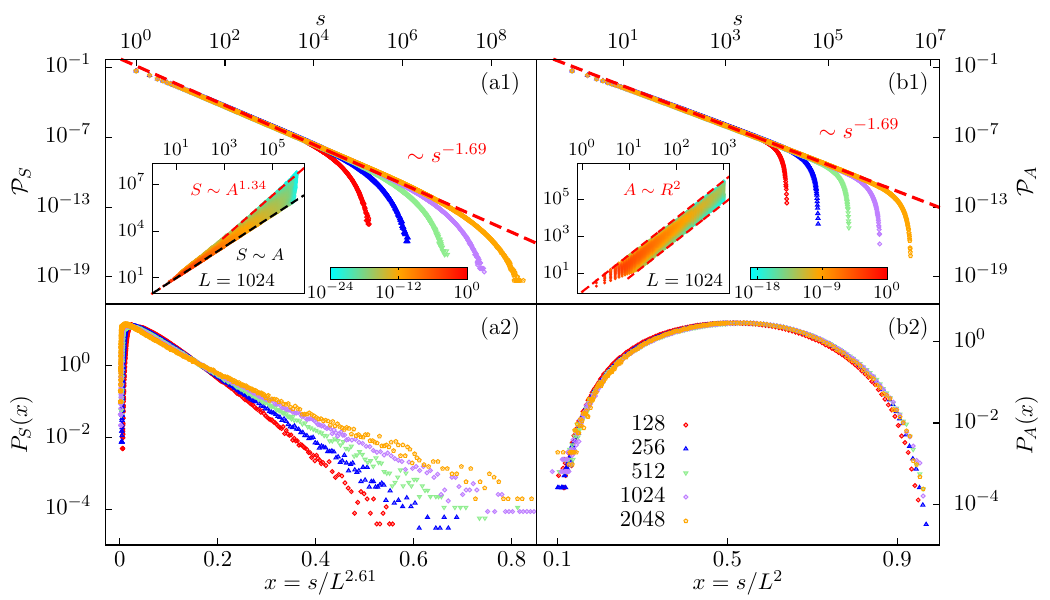}
\caption{(Color online) Self-similar dynamics in the BTW sandpile model on square lattices. Panels (a1) and (b1) show the dynamic distributions of avalanche sizes ${\cal P}_S(s,L)$ and of avalanche areas ${\cal P}_A(s,L)$, respectively, over the nonequilibrium process from the inception up to the pseudocritical time $t_L$. Both distributions exhibit clean power-law behaviors with a universal exponent $\tau' \approx 1.69$. The insets show the joint distributions $\mathcal{P}(S,A)$ and $\mathcal{P}(A,R)$, respectively, where $R$ is the avalanche diameter and the color represents probability density. Panels (a2) and (b2) present the rescaled distributions $P_S(x)$ and $P_A(x)$ at $t_L$, with $x\equiv s/L^{2.61}$ and $x\equiv s/L^2$, respectively.} \label{f5}
\end{figure}

The dynamic sampling method can also reveal nonequilibrium scaling behaviors in SOC systems. As a representative example, we consider the BTW sandpile model on a square lattice with open boundaries. Starting from an empty lattice, grains are added one by one to random sites. When a site accumulates four or more grains, it topples, redistributing grains to its four neighbors and possibly triggering further topplings. Grains can exit the system via the boundaries. After sufficiently long evolution, the system reaches a stationary critical state without any tuning parameter, characteristic of SOC.

Instead of focusing on the stationary regime as in conventional studies, we examine the initial nonequilibrium dynamics from the inception up to a pseudocritical time $t_L$, defined as the earliest avalanche that spans the system by reaching two opposite boundaries. At each time $t \leq t_L$, we record the total number of topplings ${\cal S}$ (toppled grains) and the toppling area ${\cal A}$ (distinct toppled sites), and compute their dynamic distributions $\mathcal{P}_S$ and $\mathcal{P}_A$. As shown in Figs.~\ref{f5} (a1) and (b1), both $\mathcal{P}_S$ and $\mathcal{P}_A$ exhibit power-law behavior $\sim s^{-\tau'}$, sharing the same exponent $\tau' \approx 1.69$. This value significantly differs from the stationary SOC exponents, which range from $1.05$ to $1.29$~\cite{Luebeck1997,Kadanoff1989,Manna1991,Milshtein1998,Manna2025,Engsig2025}, highlighting the distinct nature of the early-time dynamics.

We also analyze the avalanche diameter ${\cal R}$, defined as the span distance of toppled sites along one direction. The joint distribution $\mathcal{P}(A, R)$ (inset of Fig.~\ref{f5} (b1)) reveals that $\mathcal{A}$ scales as $\sim \mathcal{R}^2$, both in average and fluctuations, indicating a fractal dimension $d_A = 2$. In contrast, the distribution $\mathcal{P}(S,A)$ (inset of Fig.~\ref{f5} (a1)) broadens with increasing $A$: its lower boundary $\mathcal{S} \sim \mathcal{A}$ corresponds to avalanches with single topplings per site, while the upper boundary $\mathcal{S} \sim \mathcal{A}^\kappa$ with $\kappa \gtrsim 1.34$ captures cascades involving multiple topplings per site. If one instead considers only averaged quantities, $S=\langle \mathcal{S} \rangle$ and $A=\langle \mathcal{A} \rangle$, a simple scaling between $S$ and $A$ can still be observed, but it fails to reveal the correct underlying scaling behavior.

To further compare toppling area $\mathcal{A}$ and toppling size $\mathcal{S}$, we examine their distributions $P_A(s,L)$ and $P_S(s,L)$ at $t_L$. For toppling area, plotting the rescaled variable $x \equiv s/L^2$ leads to excellent data collapse across different $L$ (see Fig.~\ref{f5} (b2)), confirming that $A$ is a well-defined fractal object with $d_A = d$. However, for toppling size, no single exponent $d_S$ leads to full data collapse. As shown in Fig.~\ref{f5} (a2), to align the small-$s$ regime, a small exponent $d_S \approx 2$ is needed, while a larger value $d_S \approx 2.61$ matches the peaks. Furthermore, the semi-log plot in Fig.~\ref{f5} (a2) suggests that the large-$s$ tail decays exponentially as $\sim e^{-s/L^{d_S}}$, indicating a non-power-law cutoff. Thus, $S$ is not characterized by a single fractal dimension and cannot be viewed as a well-defined fractal object in the conventional sense.

In a conventional analysis for the SOC regime~\cite{Tebaldi1999}, Tebaldi et al. performed FSS of the avalanche-size moments $\langle \mathcal{S}^q \rangle \sim L^{\theta(q)}$ and found $\theta(1) = 2$, $\theta(2)-\theta(1) = 2.7$, and $\theta(9)-\theta(8) = 2.9$, suggesting that the size statistics exhibit weak multifractality. Our dynamic distributions $\mathcal{P}(S,A)$ and $\mathcal{P}_S(s,L)$ offer a more direct and vivid manifestation of this multifractal behavior. Based on the exponential cutoff form, we further conjecture that the maximal cutoff exponent for $S$ is $d_S = 3$.

\section{Discussion}    \label{sec-dis}

We reveal robust temporal self-similarity in various dynamic observables in a wide variety of percolation systems and BTW sandpile models. The self-similar dynamics not only captures scaling behaviors that were thought to emerge only in critical spatial structures, but also reveals rich critical phenomena that are challenging to be extracted by static analysis. Specifically, we theoretically established the dynamic scaling relations, $\tau_G'=\tau$ and $\tau_S'=\tau-\sigma-1$, linking the dynamic Fisher exponents directly to the static exponents.

The dynamic method serves as a powerful new paradigm for extracting critical behavior by incorporating both temporal evolution and spatial information. First, dynamic self-similarity integrates critical properties over a broad parameter range, thereby avoiding the need for precise identification of the critical point. Second, beyond temporal observables, spatial quantities sampled at the pseudocritical point $t_L$ can be analyzed using FSS theory, and the flexibility in defining $t_L$ provides an additional degree of freedom to tailor the analysis and mitigate finite-size corrections. Important applications include high-dimensional percolation~\cite{Li2024a}, long-ranged percolation~\cite{Liu2025}, and percolation on SF networks~\cite{Zhao2025}. Third, dynamic approaches can reveal critical behavior that remains elusive under static averaging. A representative example is EP, whose critical behavior has been clarified by dynamic analyses~\cite{Li2023,Li2024,Yang2024}. Fourth, dynamic analysis may encode richer information than static snapshots, such as the cascade effect in rigidity percolation, and the weak multifractal character in the BTW sandpile model.

More broadly, the dynamic formulation provides a natural framework for systems whose critical behavior emerges through continuous evolution rather than well-defined steady states. It is readily applicable to a wide range of models with complex dynamics, including jamming and cascading failures, and offers promising perspectives for studying infrastructure resilience, biological systems, and geophysical processes~\cite{Fang2024}, where static descriptions are often insufficient.

\section{Acknowledgements}

The research was supported by the National Natural Science Foundation of China under Grant No.~12275263, the Innovation Program for Quantum Science and Technology under Grant No.~2021ZD0301900, and Natural Science Foundation of Fujian province of China under Grant No.~2023J02032.

\bibliography{ref}

\begin{thebibliography}{70}%
\makeatletter
\providecommand \@ifxundefined [1]{%
 \@ifx{#1\undefined}
}%
\providecommand \@ifnum [1]{%
 \ifnum #1\expandafter \@firstoftwo
 \else \expandafter \@secondoftwo
 \fi
}%
\providecommand \@ifx [1]{%
 \ifx #1\expandafter \@firstoftwo
 \else \expandafter \@secondoftwo
 \fi
}%
\providecommand \natexlab [1]{#1}%
\providecommand \enquote  [1]{``#1''}%
\providecommand \bibnamefont  [1]{#1}%
\providecommand \bibfnamefont [1]{#1}%
\providecommand \citenamefont [1]{#1}%
\providecommand \href@noop [0]{\@secondoftwo}%
\providecommand \href [0]{\begingroup \@sanitize@url \@href}%
\providecommand \@href[1]{\@@startlink{#1}\@@href}%
\providecommand \@@href[1]{\endgroup#1\@@endlink}%
\providecommand \@sanitize@url [0]{\catcode `\\12\catcode `\$12\catcode
  `\&12\catcode `\#12\catcode `\^12\catcode `\_12\catcode `\%12\relax}%
\providecommand \@@startlink[1]{}%
\providecommand \@@endlink[0]{}%
\providecommand \url  [0]{\begingroup\@sanitize@url \@url }%
\providecommand \@url [1]{\endgroup\@href {#1}{\urlprefix }}%
\providecommand \urlprefix  [0]{URL }%
\providecommand \Eprint [0]{\href }%
\providecommand \doibase [0]{https://doi.org/}%
\providecommand \selectlanguage [0]{\@gobble}%
\providecommand \bibinfo  [0]{\@secondoftwo}%
\providecommand \bibfield  [0]{\@secondoftwo}%
\providecommand \translation [1]{[#1]}%
\providecommand \BibitemOpen [0]{}%
\providecommand \bibitemStop [0]{}%
\providecommand \bibitemNoStop [0]{.\EOS\space}%
\providecommand \EOS [0]{\spacefactor3000\relax}%
\providecommand \BibitemShut  [1]{\csname bibitem#1\endcsname}%
\let\auto@bib@innerbib\@empty
\bibitem [{\citenamefont {Ma}(2018)}]{Ma2018}%
  \BibitemOpen
  \bibfield  {author} {\bibinfo {author} {\bibfnamefont {S.-K.}\ \bibnamefont
  {Ma}},\ }\href {https://doi.org/10.4324/9780429498886} {\emph {\bibinfo
  {title} {Modern Theory Of Critical Phenomena}}}\ (\bibinfo  {publisher}
  {Routledge},\ \bibinfo {year} {2018})\BibitemShut {NoStop}%
\bibitem [{\citenamefont {Stauffer}\ and\ \citenamefont
  {Aharony}(1991)}]{Stauffer1991}%
  \BibitemOpen
  \bibfield  {author} {\bibinfo {author} {\bibfnamefont {D.}~\bibnamefont
  {Stauffer}}\ and\ \bibinfo {author} {\bibfnamefont {A.}~\bibnamefont
  {Aharony}},\ }\href@noop {} {\emph {\bibinfo {title} {Introduction to
  percolation theory}}},\ \bibinfo {edition} {2nd}\ ed.\ (\bibinfo  {publisher}
  {Taylor \& Francis},\ \bibinfo {address} {London},\ \bibinfo {year}
  {1991})\BibitemShut {NoStop}%
\bibitem [{\citenamefont {Nienhuis}\ \emph {et~al.}(1980)\citenamefont
  {Nienhuis}, \citenamefont {Riedel},\ and\ \citenamefont
  {Schick}}]{Nienhuis1980}%
  \BibitemOpen
  \bibfield  {author} {\bibinfo {author} {\bibfnamefont {B.}~\bibnamefont
  {Nienhuis}}, \bibinfo {author} {\bibfnamefont {E.~K.}\ \bibnamefont
  {Riedel}},\ and\ \bibinfo {author} {\bibfnamefont {M.}~\bibnamefont
  {Schick}},\ }\bibfield  {title} {\bibinfo {title} {Magnetic exponents of the
  two-dimensional q-state {P}otts model},\ }\href
  {https://doi.org/10.1088/0305-4470/13/6/005} {\bibfield  {journal} {\bibinfo
  {journal} {J. Phys. A: Math. Gen.}\ }\textbf {\bibinfo {volume} {13}},\
  \bibinfo {pages} {L189} (\bibinfo {year} {1980})}\BibitemShut {NoStop}%
\bibitem [{\citenamefont {Nienhuis}(1984)}]{Nienhuis1984}%
  \BibitemOpen
  \bibfield  {author} {\bibinfo {author} {\bibfnamefont {B.}~\bibnamefont
  {Nienhuis}},\ }\bibfield  {title} {\bibinfo {title} {Critical behavior of
  two-dimensional spin models and charge asymmetry in the {Coulomb} gas},\
  }\href {https://doi.org/10.1007/bf01009437} {\bibfield  {journal} {\bibinfo
  {journal} {J. Stat. Phys.}\ }\textbf {\bibinfo {volume} {34}},\ \bibinfo
  {pages} {731} (\bibinfo {year} {1984})}\BibitemShut {NoStop}%
\bibitem [{\citenamefont {Nienhuis}(1987)}]{Nienhuis1987}%
  \BibitemOpen
  \bibfield  {author} {\bibinfo {author} {\bibfnamefont {B.}~\bibnamefont
  {Nienhuis}},\ }\href@noop {} {\emph {\bibinfo {title} {Coulomb gas
  formulation of 2{D} phase transition, in ``{P}hase transition and critical
  phenomena”}}},\ edited by\ \bibinfo {editor} {\bibfnamefont
  {J.}~\bibnamefont {{C. Domb}}}\ (\bibinfo  {publisher} {Academic Press
  London},\ \bibinfo {address} {London},\ \bibinfo {year} {1987})\ \bibinfo
  {note} {vol. 11}\BibitemShut {NoStop}%
\bibitem [{\citenamefont {Cardy}(1987)}]{Cardy1987}%
  \BibitemOpen
  \bibfield  {author} {\bibinfo {author} {\bibfnamefont {J.~L.}\ \bibnamefont
  {Cardy}},\ }\href@noop {} {\emph {\bibinfo {title} {Conformal Invariance, in
  ``{P}hase transition and critical phenomena”}}},\ edited by\ \bibinfo
  {editor} {\bibfnamefont {J.}~\bibnamefont {{C. Domb}}},\ Vol.\ \bibinfo
  {volume} {11, p.55}\ (\bibinfo  {publisher} {Academic Press},\ \bibinfo
  {address} {London},\ \bibinfo {year} {1987})\BibitemShut {NoStop}%
\bibitem [{\citenamefont {Smirnov}\ and\ \citenamefont
  {Werner}(2001)}]{Smirnov2001}%
  \BibitemOpen
  \bibfield  {author} {\bibinfo {author} {\bibfnamefont {S.}~\bibnamefont
  {Smirnov}}\ and\ \bibinfo {author} {\bibfnamefont {W.}~\bibnamefont
  {Werner}},\ }\bibfield  {title} {\bibinfo {title} {{Critical exponents for
  two-dimensional percolation}},\ }\href {https://hal.science/hal-00119178}
  {\bibfield  {journal} {\bibinfo  {journal} {Math. Res. Lett.}\ }\textbf
  {\bibinfo {volume} {8}},\ \bibinfo {pages} {729} (\bibinfo {year}
  {2001})}\BibitemShut {NoStop}%
\bibitem [{\citenamefont {Achlioptas}\ \emph {et~al.}(2009)\citenamefont
  {Achlioptas}, \citenamefont {D’Souza},\ and\ \citenamefont
  {Spencer}}]{Achlioptas2009}%
  \BibitemOpen
  \bibfield  {author} {\bibinfo {author} {\bibfnamefont {D.}~\bibnamefont
  {Achlioptas}}, \bibinfo {author} {\bibfnamefont {R.~M.}\ \bibnamefont
  {D’Souza}},\ and\ \bibinfo {author} {\bibfnamefont {J.}~\bibnamefont
  {Spencer}},\ }\bibfield  {title} {\bibinfo {title} {Explosive percolation in
  random networks},\ }\href {https://doi.org/10.1126/science.1167782}
  {\bibfield  {journal} {\bibinfo  {journal} {Science}\ }\textbf {\bibinfo
  {volume} {323}},\ \bibinfo {pages} {1453} (\bibinfo {year}
  {2009})}\BibitemShut {NoStop}%
\bibitem [{\citenamefont {Riordan}\ and\ \citenamefont
  {Warnke}(2011)}]{Riordan2011}%
  \BibitemOpen
  \bibfield  {author} {\bibinfo {author} {\bibfnamefont {O.}~\bibnamefont
  {Riordan}}\ and\ \bibinfo {author} {\bibfnamefont {L.}~\bibnamefont
  {Warnke}},\ }\bibfield  {title} {\bibinfo {title} {Explosive percolation is
  continuous},\ }\href {https://doi.org/10.1126/science.1206241} {\bibfield
  {journal} {\bibinfo  {journal} {Science}\ }\textbf {\bibinfo {volume}
  {333}},\ \bibinfo {pages} {322} (\bibinfo {year} {2011})}\BibitemShut
  {NoStop}%
\bibitem [{\citenamefont {Jacobs}\ and\ \citenamefont
  {Thorpe}(1995)}]{Jacobs1995}%
  \BibitemOpen
  \bibfield  {author} {\bibinfo {author} {\bibfnamefont {D.~J.}\ \bibnamefont
  {Jacobs}}\ and\ \bibinfo {author} {\bibfnamefont {M.~F.}\ \bibnamefont
  {Thorpe}},\ }\bibfield  {title} {\bibinfo {title} {Generic rigidity
  percolation: The pebble game},\ }\href
  {https://doi.org/10.1103/PhysRevLett.75.4051} {\bibfield  {journal} {\bibinfo
   {journal} {Phys. Rev. Lett.}\ }\textbf {\bibinfo {volume} {75}},\ \bibinfo
  {pages} {4051} (\bibinfo {year} {1995})}\BibitemShut {NoStop}%
\bibitem [{\citenamefont {Jacobs}\ and\ \citenamefont
  {Thorpe}(1996)}]{Jacobs1996}%
  \BibitemOpen
  \bibfield  {author} {\bibinfo {author} {\bibfnamefont {D.~J.}\ \bibnamefont
  {Jacobs}}\ and\ \bibinfo {author} {\bibfnamefont {M.~F.}\ \bibnamefont
  {Thorpe}},\ }\bibfield  {title} {\bibinfo {title} {Generic rigidity
  percolation in two dimensions},\ }\href
  {https://doi.org/10.1103/PhysRevE.53.3682} {\bibfield  {journal} {\bibinfo
  {journal} {Phys. Rev. E}\ }\textbf {\bibinfo {volume} {53}},\ \bibinfo
  {pages} {3682} (\bibinfo {year} {1996})}\BibitemShut {NoStop}%
\bibitem [{\citenamefont {Moukarzel}\ and\ \citenamefont
  {Duxbury}(1999)}]{Moukarzel1999}%
  \BibitemOpen
  \bibfield  {author} {\bibinfo {author} {\bibfnamefont {C.}~\bibnamefont
  {Moukarzel}}\ and\ \bibinfo {author} {\bibfnamefont {P.~M.}\ \bibnamefont
  {Duxbury}},\ }\bibfield  {title} {\bibinfo {title} {Comparison of rigidity
  and connectivity percolation in two dimensions},\ }\href
  {https://doi.org/10.1103/PhysRevE.59.2614} {\bibfield  {journal} {\bibinfo
  {journal} {Phys. Rev. E}\ }\textbf {\bibinfo {volume} {59}},\ \bibinfo
  {pages} {2614} (\bibinfo {year} {1999})}\BibitemShut {NoStop}%
\bibitem [{\citenamefont {Bri\`ere}\ \emph {et~al.}(2007)\citenamefont
  {Bri\`ere}, \citenamefont {Chubynsky},\ and\ \citenamefont
  {Mousseau}}]{Briere2007}%
  \BibitemOpen
  \bibfield  {author} {\bibinfo {author} {\bibfnamefont {M.-A.}\ \bibnamefont
  {Bri\`ere}}, \bibinfo {author} {\bibfnamefont {M.~V.}\ \bibnamefont
  {Chubynsky}},\ and\ \bibinfo {author} {\bibfnamefont {N.}~\bibnamefont
  {Mousseau}},\ }\bibfield  {title} {\bibinfo {title} {Self-organized
  criticality in the intermediate phase of rigidity percolation},\ }\href
  {https://doi.org/10.1103/PhysRevE.75.056108} {\bibfield  {journal} {\bibinfo
  {journal} {Phys. Rev. E}\ }\textbf {\bibinfo {volume} {75}},\ \bibinfo
  {pages} {056108} (\bibinfo {year} {2007})}\BibitemShut {NoStop}%
\bibitem [{\citenamefont {Friedman}\ and\ \citenamefont
  {Landsberg}(2009)}]{Friedman2009}%
  \BibitemOpen
  \bibfield  {author} {\bibinfo {author} {\bibfnamefont {E.~J.}\ \bibnamefont
  {Friedman}}\ and\ \bibinfo {author} {\bibfnamefont {A.~S.}\ \bibnamefont
  {Landsberg}},\ }\bibfield  {title} {\bibinfo {title} {Construction and
  analysis of random networks with explosive percolation},\ }\href
  {https://doi.org/10.1103/physrevlett.103.255701} {\bibfield  {journal}
  {\bibinfo  {journal} {Phys. Rev. Lett.}\ }\textbf {\bibinfo {volume} {103}},\
  \bibinfo {pages} {255701} (\bibinfo {year} {2009})}\BibitemShut {NoStop}%
\bibitem [{\citenamefont {Ziff}(2009)}]{Ziff2009}%
  \BibitemOpen
  \bibfield  {author} {\bibinfo {author} {\bibfnamefont {R.~M.}\ \bibnamefont
  {Ziff}},\ }\bibfield  {title} {\bibinfo {title} {Explosive growth in biased
  dynamic percolation on two-dimensional regular lattice networks},\ }\href
  {https://doi.org/10.1103/physrevlett.103.045701} {\bibfield  {journal}
  {\bibinfo  {journal} {Phys. Rev. Lett.}\ }\textbf {\bibinfo {volume} {103}},\
  \bibinfo {pages} {045701} (\bibinfo {year} {2009})}\BibitemShut {NoStop}%
\bibitem [{\citenamefont {D’Souza}\ and\ \citenamefont
  {Mitzenmacher}(2010)}]{Souza2010}%
  \BibitemOpen
  \bibfield  {author} {\bibinfo {author} {\bibfnamefont {R.~M.}\ \bibnamefont
  {D’Souza}}\ and\ \bibinfo {author} {\bibfnamefont {M.}~\bibnamefont
  {Mitzenmacher}},\ }\bibfield  {title} {\bibinfo {title} {Local cluster
  aggregation models of explosive percolation},\ }\href
  {https://doi.org/10.1103/physrevlett.104.195702} {\bibfield  {journal}
  {\bibinfo  {journal} {Phys. Rev. Lett.}\ }\textbf {\bibinfo {volume} {104}},\
  \bibinfo {pages} {195702} (\bibinfo {year} {2010})}\BibitemShut {NoStop}%
\bibitem [{\citenamefont {Radicchi}\ and\ \citenamefont
  {Fortunato}(2010)}]{Radicchi2010}%
  \BibitemOpen
  \bibfield  {author} {\bibinfo {author} {\bibfnamefont {F.}~\bibnamefont
  {Radicchi}}\ and\ \bibinfo {author} {\bibfnamefont {S.}~\bibnamefont
  {Fortunato}},\ }\bibfield  {title} {\bibinfo {title} {Explosive percolation:
  A numerical analysis},\ }\href {https://doi.org/10.1103/physreve.81.036110}
  {\bibfield  {journal} {\bibinfo  {journal} {Phys. Rev. E}\ }\textbf {\bibinfo
  {volume} {81}},\ \bibinfo {pages} {036110} (\bibinfo {year}
  {2010})}\BibitemShut {NoStop}%
\bibitem [{\citenamefont {Grassberger}\ \emph {et~al.}(2011)\citenamefont
  {Grassberger}, \citenamefont {Christensen}, \citenamefont {Bizhani},
  \citenamefont {Son},\ and\ \citenamefont {Paczuski}}]{Grassberger2011}%
  \BibitemOpen
  \bibfield  {author} {\bibinfo {author} {\bibfnamefont {P.}~\bibnamefont
  {Grassberger}}, \bibinfo {author} {\bibfnamefont {C.}~\bibnamefont
  {Christensen}}, \bibinfo {author} {\bibfnamefont {G.}~\bibnamefont
  {Bizhani}}, \bibinfo {author} {\bibfnamefont {S.-W.}\ \bibnamefont {Son}},\
  and\ \bibinfo {author} {\bibfnamefont {M.}~\bibnamefont {Paczuski}},\
  }\bibfield  {title} {\bibinfo {title} {Explosive percolation is continuous,
  but with unusual finite size behavior},\ }\href
  {https://doi.org/10.1103/physrevlett.106.225701} {\bibfield  {journal}
  {\bibinfo  {journal} {Phys. Rev. Lett.}\ }\textbf {\bibinfo {volume} {106}},\
  \bibinfo {pages} {225701} (\bibinfo {year} {2011})}\BibitemShut {NoStop}%
\bibitem [{\citenamefont {D’Souza}\ and\ \citenamefont
  {Nagler}(2015)}]{Souza2015}%
  \BibitemOpen
  \bibfield  {author} {\bibinfo {author} {\bibfnamefont {R.~M.}\ \bibnamefont
  {D’Souza}}\ and\ \bibinfo {author} {\bibfnamefont {J.}~\bibnamefont
  {Nagler}},\ }\bibfield  {title} {\bibinfo {title} {Anomalous critical and
  supercritical phenomena in explosive percolation},\ }\href
  {https://doi.org/10.1038/nphys3378} {\bibfield  {journal} {\bibinfo
  {journal} {Nat. Phys.}\ }\textbf {\bibinfo {volume} {11}},\ \bibinfo {pages}
  {531} (\bibinfo {year} {2015})}\BibitemShut {NoStop}%
\bibitem [{\citenamefont {Feng}\ and\ \citenamefont {Sen}(1984)}]{Feng1984}%
  \BibitemOpen
  \bibfield  {author} {\bibinfo {author} {\bibfnamefont {S.}~\bibnamefont
  {Feng}}\ and\ \bibinfo {author} {\bibfnamefont {P.~N.}\ \bibnamefont {Sen}},\
  }\bibfield  {title} {\bibinfo {title} {Percolation on elastic networks: New
  exponent and threshold},\ }\href {https://doi.org/10.1103/physrevlett.52.216}
  {\bibfield  {journal} {\bibinfo  {journal} {Phys. Rev. Lett.}\ }\textbf
  {\bibinfo {volume} {52}},\ \bibinfo {pages} {216} (\bibinfo {year}
  {1984})}\BibitemShut {NoStop}%
\bibitem [{\citenamefont {Kantor}\ and\ \citenamefont
  {Webman}(1984)}]{Kantor1984}%
  \BibitemOpen
  \bibfield  {author} {\bibinfo {author} {\bibfnamefont {Y.}~\bibnamefont
  {Kantor}}\ and\ \bibinfo {author} {\bibfnamefont {I.}~\bibnamefont
  {Webman}},\ }\bibfield  {title} {\bibinfo {title} {Elastic properties of
  random percolating systems},\ }\href
  {https://doi.org/10.1103/physrevlett.52.1891} {\bibfield  {journal} {\bibinfo
   {journal} {Phys. Rev. Lett.}\ }\textbf {\bibinfo {volume} {52}},\ \bibinfo
  {pages} {1891} (\bibinfo {year} {1984})}\BibitemShut {NoStop}%
\bibitem [{\citenamefont {Hansen}\ and\ \citenamefont
  {Roux}(1989)}]{Hansen1989}%
  \BibitemOpen
  \bibfield  {author} {\bibinfo {author} {\bibfnamefont {A.}~\bibnamefont
  {Hansen}}\ and\ \bibinfo {author} {\bibfnamefont {S.}~\bibnamefont {Roux}},\
  }\bibfield  {title} {\bibinfo {title} {Universality class of central-force
  percolation},\ }\href {https://doi.org/10.1103/PhysRevB.40.749} {\bibfield
  {journal} {\bibinfo  {journal} {Phys. Rev. B}\ }\textbf {\bibinfo {volume}
  {40}},\ \bibinfo {pages} {749} (\bibinfo {year} {1989})}\BibitemShut
  {NoStop}%
\bibitem [{\citenamefont {Plischke}(2007)}]{Plischke2007}%
  \BibitemOpen
  \bibfield  {author} {\bibinfo {author} {\bibfnamefont {M.}~\bibnamefont
  {Plischke}},\ }\bibfield  {title} {\bibinfo {title} {Rigidity of disordered
  networks with bond-bending forces},\ }\href
  {https://doi.org/10.1103/physreve.76.021401} {\bibfield  {journal} {\bibinfo
  {journal} {Phys. Rev. E}\ }\textbf {\bibinfo {volume} {76}},\ \bibinfo
  {pages} {021401} (\bibinfo {year} {2007})}\BibitemShut {NoStop}%
\bibitem [{\citenamefont {Zhang}\ \emph {et~al.}(2015)\citenamefont {Zhang},
  \citenamefont {Rocklin}, \citenamefont {Chen},\ and\ \citenamefont
  {Mao}}]{Zhang2015}%
  \BibitemOpen
  \bibfield  {author} {\bibinfo {author} {\bibfnamefont {L.}~\bibnamefont
  {Zhang}}, \bibinfo {author} {\bibfnamefont {D.~Z.}\ \bibnamefont {Rocklin}},
  \bibinfo {author} {\bibfnamefont {B.~G.-g.}\ \bibnamefont {Chen}},\ and\
  \bibinfo {author} {\bibfnamefont {X.}~\bibnamefont {Mao}},\ }\bibfield
  {title} {\bibinfo {title} {Rigidity percolation by next-nearest-neighbor
  bonds on generic and regular isostatic lattices},\ }\href
  {https://doi.org/10.1103/physreve.91.032124} {\bibfield  {journal} {\bibinfo
  {journal} {Phys. Rev. E}\ }\textbf {\bibinfo {volume} {91}},\ \bibinfo
  {pages} {032124} (\bibinfo {year} {2015})}\BibitemShut {NoStop}%
\bibitem [{\citenamefont {Ellenbroek}\ \emph {et~al.}(2015)\citenamefont
  {Ellenbroek}, \citenamefont {Hagh}, \citenamefont {Kumar}, \citenamefont
  {Thorpe},\ and\ \citenamefont {van Hecke}}]{Ellenbroek2015}%
  \BibitemOpen
  \bibfield  {author} {\bibinfo {author} {\bibfnamefont {W.~G.}\ \bibnamefont
  {Ellenbroek}}, \bibinfo {author} {\bibfnamefont {V.~F.}\ \bibnamefont
  {Hagh}}, \bibinfo {author} {\bibfnamefont {A.}~\bibnamefont {Kumar}},
  \bibinfo {author} {\bibfnamefont {M.}~\bibnamefont {Thorpe}},\ and\ \bibinfo
  {author} {\bibfnamefont {M.}~\bibnamefont {van Hecke}},\ }\bibfield  {title}
  {\bibinfo {title} {Rigidity loss in disordered systems: Three scenarios},\
  }\href {https://doi.org/10.1103/physrevlett.114.135501} {\bibfield  {journal}
  {\bibinfo  {journal} {Phys. Rev. Lett.}\ }\textbf {\bibinfo {volume} {114}},\
  \bibinfo {pages} {135501} (\bibinfo {year} {2015})}\BibitemShut {NoStop}%
\bibitem [{\citenamefont {Lu}\ \emph {et~al.}(2024)\citenamefont {Lu},
  \citenamefont {Song}, \citenamefont {Li},\ and\ \citenamefont
  {Deng}}]{Lu2024}%
  \BibitemOpen
  \bibfield  {author} {\bibinfo {author} {\bibfnamefont {M.}~\bibnamefont
  {Lu}}, \bibinfo {author} {\bibfnamefont {Y.-F.}\ \bibnamefont {Song}},
  \bibinfo {author} {\bibfnamefont {M.}~\bibnamefont {Li}},\ and\ \bibinfo
  {author} {\bibfnamefont {Y.}~\bibnamefont {Deng}},\ }\bibfield  {title}
  {\bibinfo {title} {Self-similar gap dynamics in percolation and rigidity
  percolation},\ }\href {https://arxiv.org/abs/2411.04748} {\bibfield
  {journal} {\bibinfo  {journal} {arXiv preprint}\ ,\ \bibinfo {pages}
  {2411.04748}} (\bibinfo {year} {2024})}\BibitemShut {NoStop}%
\bibitem [{\citenamefont {Buldyrev}\ \emph {et~al.}(2010)\citenamefont
  {Buldyrev}, \citenamefont {Parshani}, \citenamefont {Paul}, \citenamefont
  {Stanley},\ and\ \citenamefont {Havlin}}]{Buldyrev2010}%
  \BibitemOpen
  \bibfield  {author} {\bibinfo {author} {\bibfnamefont {S.~V.}\ \bibnamefont
  {Buldyrev}}, \bibinfo {author} {\bibfnamefont {R.}~\bibnamefont {Parshani}},
  \bibinfo {author} {\bibfnamefont {G.}~\bibnamefont {Paul}}, \bibinfo {author}
  {\bibfnamefont {H.~E.}\ \bibnamefont {Stanley}},\ and\ \bibinfo {author}
  {\bibfnamefont {S.}~\bibnamefont {Havlin}},\ }\bibfield  {title} {\bibinfo
  {title} {Catastrophic cascade of failures in interdependent networks},\
  }\href {https://doi.org/10.1038/nature08932} {\bibfield  {journal} {\bibinfo
  {journal} {Nature}\ }\textbf {\bibinfo {volume} {464}},\ \bibinfo {pages}
  {1025} (\bibinfo {year} {2010})}\BibitemShut {NoStop}%
\bibitem [{\citenamefont {Fan}\ \emph {et~al.}(2018)\citenamefont {Fan},
  \citenamefont {Meng}, \citenamefont {Ashkenazy}, \citenamefont {Havlin},\
  and\ \citenamefont {Schellnhuber}}]{Fan2018}%
  \BibitemOpen
  \bibfield  {author} {\bibinfo {author} {\bibfnamefont {J.}~\bibnamefont
  {Fan}}, \bibinfo {author} {\bibfnamefont {J.}~\bibnamefont {Meng}}, \bibinfo
  {author} {\bibfnamefont {Y.}~\bibnamefont {Ashkenazy}}, \bibinfo {author}
  {\bibfnamefont {S.}~\bibnamefont {Havlin}},\ and\ \bibinfo {author}
  {\bibfnamefont {H.~J.}\ \bibnamefont {Schellnhuber}},\ }\bibfield  {title}
  {\bibinfo {title} {Climate network percolation reveals the expansion and
  weakening of the tropical component under global warming},\ }\href
  {https://doi.org/10.1073/pnas.1811068115} {\bibfield  {journal} {\bibinfo
  {journal} {Proc. Natl. Acad. Sci.}\ }\textbf {\bibinfo {volume} {115}},\
  \bibinfo {pages} {E12128} (\bibinfo {year} {2018})}\BibitemShut {NoStop}%
\bibitem [{\citenamefont {Li}\ \emph {et~al.}(2014)\citenamefont {Li},
  \citenamefont {Fu}, \citenamefont {Wang}, \citenamefont {Lu}, \citenamefont
  {Berezin}, \citenamefont {Stanley},\ and\ \citenamefont {Havlin}}]{Li2014}%
  \BibitemOpen
  \bibfield  {author} {\bibinfo {author} {\bibfnamefont {D.}~\bibnamefont
  {Li}}, \bibinfo {author} {\bibfnamefont {B.}~\bibnamefont {Fu}}, \bibinfo
  {author} {\bibfnamefont {Y.}~\bibnamefont {Wang}}, \bibinfo {author}
  {\bibfnamefont {G.}~\bibnamefont {Lu}}, \bibinfo {author} {\bibfnamefont
  {Y.}~\bibnamefont {Berezin}}, \bibinfo {author} {\bibfnamefont {H.~E.}\
  \bibnamefont {Stanley}},\ and\ \bibinfo {author} {\bibfnamefont
  {S.}~\bibnamefont {Havlin}},\ }\bibfield  {title} {\bibinfo {title}
  {Percolation transition in dynamical traffic network with evolving critical
  bottlenecks},\ }\href {https://doi.org/10.1073/pnas.1419185112} {\bibfield
  {journal} {\bibinfo  {journal} {Proc. Natl. Acad. Sci.}\ }\textbf {\bibinfo
  {volume} {112}},\ \bibinfo {pages} {669} (\bibinfo {year}
  {2014})}\BibitemShut {NoStop}%
\bibitem [{\citenamefont {Xie}\ \emph {et~al.}(2021)\citenamefont {Xie},
  \citenamefont {Meng}, \citenamefont {Sun}, \citenamefont {Ma}, \citenamefont
  {Yan},\ and\ \citenamefont {Hu}}]{Xie2021}%
  \BibitemOpen
  \bibfield  {author} {\bibinfo {author} {\bibfnamefont {J.}~\bibnamefont
  {Xie}}, \bibinfo {author} {\bibfnamefont {F.}~\bibnamefont {Meng}}, \bibinfo
  {author} {\bibfnamefont {J.}~\bibnamefont {Sun}}, \bibinfo {author}
  {\bibfnamefont {X.}~\bibnamefont {Ma}}, \bibinfo {author} {\bibfnamefont
  {G.}~\bibnamefont {Yan}},\ and\ \bibinfo {author} {\bibfnamefont
  {Y.}~\bibnamefont {Hu}},\ }\bibfield  {title} {\bibinfo {title} {Detecting
  and modelling real percolation and phase transitions of information on social
  media},\ }\href {https://doi.org/10.1038/s41562-021-01090-z} {\bibfield
  {journal} {\bibinfo  {journal} {Nat. Hum. Behav.}\ }\textbf {\bibinfo
  {volume} {5}},\ \bibinfo {pages} {1161} (\bibinfo {year} {2021})}\BibitemShut
  {NoStop}%
\bibitem [{\citenamefont {Lu}\ \emph {et~al.}(2026)\citenamefont {Lu},
  \citenamefont {Song}, \citenamefont {Shi}, \citenamefont {Li},\ and\
  \citenamefont {Deng}}]{Lu2026}%
  \BibitemOpen
  \bibfield  {author} {\bibinfo {author} {\bibfnamefont {M.}~\bibnamefont
  {Lu}}, \bibinfo {author} {\bibfnamefont {Y.-F.}\ \bibnamefont {Song}},
  \bibinfo {author} {\bibfnamefont {Q.}~\bibnamefont {Shi}}, \bibinfo {author}
  {\bibfnamefont {M.}~\bibnamefont {Li}},\ and\ \bibinfo {author}
  {\bibfnamefont {Y.}~\bibnamefont {Deng}},\ }\bibfield  {title} {\bibinfo
  {title} {High-precision dynamic {Monte Carlo} study of rigidity percolaion},\
  }\href {https://doi.org/10.48550/arXiv.2601.21399} {\bibfield  {journal}
  {\bibinfo  {journal} {arXiv preprint}\ ,\ \bibinfo {pages} {2601.21399}}
  (\bibinfo {year} {2026})}\BibitemShut {NoStop}%
\bibitem [{\citenamefont {Bak}\ \emph {et~al.}(1987)\citenamefont {Bak},
  \citenamefont {Tang},\ and\ \citenamefont {Wiesenfeld}}]{Bak1987}%
  \BibitemOpen
  \bibfield  {author} {\bibinfo {author} {\bibfnamefont {P.}~\bibnamefont
  {Bak}}, \bibinfo {author} {\bibfnamefont {C.}~\bibnamefont {Tang}},\ and\
  \bibinfo {author} {\bibfnamefont {K.}~\bibnamefont {Wiesenfeld}},\ }\bibfield
   {title} {\bibinfo {title} {Self-organized criticality: An explanation of the
  1/f noise},\ }\href {https://doi.org/10.1103/physrevlett.59.381} {\bibfield
  {journal} {\bibinfo  {journal} {Phys. Rev. Lett.}\ }\textbf {\bibinfo
  {volume} {59}},\ \bibinfo {pages} {381} (\bibinfo {year} {1987})}\BibitemShut
  {NoStop}%
\bibitem [{\citenamefont {Tang}\ and\ \citenamefont {Bak}(1988)}]{Tang1988}%
  \BibitemOpen
  \bibfield  {author} {\bibinfo {author} {\bibfnamefont {C.}~\bibnamefont
  {Tang}}\ and\ \bibinfo {author} {\bibfnamefont {P.}~\bibnamefont {Bak}},\
  }\bibfield  {title} {\bibinfo {title} {Critical exponents and scaling
  relations for self-organized critical phenomena},\ }\href
  {https://doi.org/10.1103/physrevlett.60.2347} {\bibfield  {journal} {\bibinfo
   {journal} {Phys. Rev. Lett.}\ }\textbf {\bibinfo {volume} {60}},\ \bibinfo
  {pages} {2347} (\bibinfo {year} {1988})}\BibitemShut {NoStop}%
\bibitem [{\citenamefont {Engsig}\ and\ \citenamefont
  {Sneppen}(2025)}]{Engsig2025}%
  \BibitemOpen
  \bibfield  {author} {\bibinfo {author} {\bibfnamefont {M.}~\bibnamefont
  {Engsig}}\ and\ \bibinfo {author} {\bibfnamefont {K.}~\bibnamefont
  {Sneppen}},\ }\bibfield  {title} {\bibinfo {title} {Fractals in the critical
  attractor of the classical sandpile model},\ }\href
  {https://doi.org/10.1103/physrevlett.134.187201} {\bibfield  {journal}
  {\bibinfo  {journal} {Phys. Rev. Lett.}\ }\textbf {\bibinfo {volume} {134}},\
  \bibinfo {pages} {187201} (\bibinfo {year} {2025})}\BibitemShut {NoStop}%
\bibitem [{\citenamefont {Manna}(2025)}]{Manna2025}%
  \BibitemOpen
  \bibfield  {author} {\bibinfo {author} {\bibfnamefont {S.~S.}\ \bibnamefont
  {Manna}},\ }\bibfield  {title} {\bibinfo {title} {Describing self-organized
  criticality as a continuous phase transition},\ }\href
  {https://doi.org/10.1103/PhysRevE.111.024111} {\bibfield  {journal} {\bibinfo
   {journal} {Phys. Rev. E}\ }\textbf {\bibinfo {volume} {111}},\ \bibinfo
  {pages} {024111} (\bibinfo {year} {2025})}\BibitemShut {NoStop}%
\bibitem [{\citenamefont {Tebaldi}\ \emph {et~al.}(1999)\citenamefont
  {Tebaldi}, \citenamefont {De~Menech},\ and\ \citenamefont
  {Stella}}]{Tebaldi1999}%
  \BibitemOpen
  \bibfield  {author} {\bibinfo {author} {\bibfnamefont {C.}~\bibnamefont
  {Tebaldi}}, \bibinfo {author} {\bibfnamefont {M.}~\bibnamefont {De~Menech}},\
  and\ \bibinfo {author} {\bibfnamefont {A.~L.}\ \bibnamefont {Stella}},\
  }\bibfield  {title} {\bibinfo {title} {Multifractal scaling in the
  {B}ak-{T}ang-{W}iesenfeld sandpile and edge events},\ }\href
  {https://doi.org/10.1103/physrevlett.83.3952} {\bibfield  {journal} {\bibinfo
   {journal} {Phys. Rev. Lett.}\ }\textbf {\bibinfo {volume} {83}},\ \bibinfo
  {pages} {3952} (\bibinfo {year} {1999})}\BibitemShut {NoStop}%
\bibitem [{\citenamefont {Newman}\ and\ \citenamefont
  {Ziff}(2001)}]{Newman2001a}%
  \BibitemOpen
  \bibfield  {author} {\bibinfo {author} {\bibfnamefont {M.~E.~J.}\
  \bibnamefont {Newman}}\ and\ \bibinfo {author} {\bibfnamefont {R.~M.}\
  \bibnamefont {Ziff}},\ }\bibfield  {title} {\bibinfo {title} {Fast {M}onte
  {C}arlo algorithm for site or bond percolation},\ }\href
  {https://doi.org/10.1103/physreve.64.016706} {\bibfield  {journal} {\bibinfo
  {journal} {Phys. Rev. E}\ }\textbf {\bibinfo {volume} {64}},\ \bibinfo
  {pages} {016706} (\bibinfo {year} {2001})}\BibitemShut {NoStop}%
\bibitem [{\citenamefont {Dorogovtsev}\ \emph {et~al.}(2001)\citenamefont
  {Dorogovtsev}, \citenamefont {Mendes},\ and\ \citenamefont
  {Samukhin}}]{Dorogovtsev2001}%
  \BibitemOpen
  \bibfield  {author} {\bibinfo {author} {\bibfnamefont {S.~N.}\ \bibnamefont
  {Dorogovtsev}}, \bibinfo {author} {\bibfnamefont {J.~F.~F.}\ \bibnamefont
  {Mendes}},\ and\ \bibinfo {author} {\bibfnamefont {A.~N.}\ \bibnamefont
  {Samukhin}},\ }\bibfield  {title} {\bibinfo {title} {Anomalous percolation
  properties of growing networks},\ }\href
  {https://doi.org/10.1103/physreve.64.066110} {\bibfield  {journal} {\bibinfo
  {journal} {Phys. Rev. E}\ }\textbf {\bibinfo {volume} {64}},\ \bibinfo
  {pages} {066110} (\bibinfo {year} {2001})}\BibitemShut {NoStop}%
\bibitem [{\citenamefont {Bollobás}\ \emph {et~al.}(2004)\citenamefont
  {Bollobás}, \citenamefont {Janson},\ and\ \citenamefont
  {Riordan}}]{Bollobas2004}%
  \BibitemOpen
  \bibfield  {author} {\bibinfo {author} {\bibfnamefont {B.}~\bibnamefont
  {Bollobás}}, \bibinfo {author} {\bibfnamefont {S.}~\bibnamefont {Janson}},\
  and\ \bibinfo {author} {\bibfnamefont {O.}~\bibnamefont {Riordan}},\
  }\bibfield  {title} {\bibinfo {title} {The phase transition in the uniformly
  grown random graph has infinite order},\ }\href
  {https://doi.org/10.1002/rsa.20041} {\bibfield  {journal} {\bibinfo
  {journal} {Random Struct. Algor.}\ }\textbf {\bibinfo {volume} {26}},\
  \bibinfo {pages} {1} (\bibinfo {year} {2004})}\BibitemShut {NoStop}%
\bibitem [{\citenamefont {Li}\ \emph {et~al.}(2023)\citenamefont {Li},
  \citenamefont {Wang},\ and\ \citenamefont {Deng}}]{Li2023}%
  \BibitemOpen
  \bibfield  {author} {\bibinfo {author} {\bibfnamefont {M.}~\bibnamefont
  {Li}}, \bibinfo {author} {\bibfnamefont {J.}~\bibnamefont {Wang}},\ and\
  \bibinfo {author} {\bibfnamefont {Y.}~\bibnamefont {Deng}},\ }\bibfield
  {title} {\bibinfo {title} {Explosive percolation obeys standard finite-size
  scaling in an event-based ensemble},\ }\href
  {https://doi.org/10.1103/physrevlett.130.147101} {\bibfield  {journal}
  {\bibinfo  {journal} {Phys. Rev. Lett.}\ }\textbf {\bibinfo {volume} {130}},\
  \bibinfo {pages} {147101} (\bibinfo {year} {2023})}\BibitemShut {NoStop}%
\bibitem [{\citenamefont {Li}\ \emph {et~al.}(2024{\natexlab{a}})\citenamefont
  {Li}, \citenamefont {Wang},\ and\ \citenamefont {Deng}}]{Li2024}%
  \BibitemOpen
  \bibfield  {author} {\bibinfo {author} {\bibfnamefont {M.}~\bibnamefont
  {Li}}, \bibinfo {author} {\bibfnamefont {J.}~\bibnamefont {Wang}},\ and\
  \bibinfo {author} {\bibfnamefont {Y.}~\bibnamefont {Deng}},\ }\bibfield
  {title} {\bibinfo {title} {Explosive percolation in finite dimensions},\
  }\href {https://doi.org/10.1103/physrevresearch.6.033319} {\bibfield
  {journal} {\bibinfo  {journal} {Phys. Rev. Res.}\ }\textbf {\bibinfo {volume}
  {6}},\ \bibinfo {pages} {033319} (\bibinfo {year}
  {2024}{\natexlab{a}})}\BibitemShut {NoStop}%
\bibitem [{\citenamefont {Li}\ \emph {et~al.}(2024{\natexlab{b}})\citenamefont
  {Li}, \citenamefont {Fang}, \citenamefont {Fan},\ and\ \citenamefont
  {Deng}}]{Li2024a}%
  \BibitemOpen
  \bibfield  {author} {\bibinfo {author} {\bibfnamefont {M.}~\bibnamefont
  {Li}}, \bibinfo {author} {\bibfnamefont {S.}~\bibnamefont {Fang}}, \bibinfo
  {author} {\bibfnamefont {J.}~\bibnamefont {Fan}},\ and\ \bibinfo {author}
  {\bibfnamefont {Y.}~\bibnamefont {Deng}},\ }\bibfield  {title} {\bibinfo
  {title} {Crossover finite-size scaling theory and its applications in
  percolation},\ }\href {https://arxiv.org/abs/2412.06228} {\bibfield
  {journal} {\bibinfo  {journal} {arXiv preprint}\ ,\ \bibinfo {pages}
  {2412.06228}} (\bibinfo {year} {2024}{\natexlab{b}})}\BibitemShut {NoStop}%
\bibitem [{\citenamefont {Fan}\ \emph {et~al.}(2020)\citenamefont {Fan},
  \citenamefont {Meng}, \citenamefont {Liu}, \citenamefont {Saberi},
  \citenamefont {Kurths},\ and\ \citenamefont {Nagler}}]{Fan2020}%
  \BibitemOpen
  \bibfield  {author} {\bibinfo {author} {\bibfnamefont {J.}~\bibnamefont
  {Fan}}, \bibinfo {author} {\bibfnamefont {J.}~\bibnamefont {Meng}}, \bibinfo
  {author} {\bibfnamefont {Y.}~\bibnamefont {Liu}}, \bibinfo {author}
  {\bibfnamefont {A.~A.}\ \bibnamefont {Saberi}}, \bibinfo {author}
  {\bibfnamefont {J.}~\bibnamefont {Kurths}},\ and\ \bibinfo {author}
  {\bibfnamefont {J.}~\bibnamefont {Nagler}},\ }\bibfield  {title} {\bibinfo
  {title} {Universal gap scaling in percolation},\ }\href
  {https://doi.org/10.1038/s41567-019-0783-2} {\bibfield  {journal} {\bibinfo
  {journal} {Nat. Phys.}\ }\textbf {\bibinfo {volume} {16}},\ \bibinfo {pages}
  {455} (\bibinfo {year} {2020})}\BibitemShut {NoStop}%
\bibitem [{\citenamefont {Deng}\ \emph {et~al.}(2010)\citenamefont {Deng},
  \citenamefont {Zhang}, \citenamefont {Garoni}, \citenamefont {Sokal},\ and\
  \citenamefont {Sportiello}}]{Deng2010}%
  \BibitemOpen
  \bibfield  {author} {\bibinfo {author} {\bibfnamefont {Y.}~\bibnamefont
  {Deng}}, \bibinfo {author} {\bibfnamefont {W.}~\bibnamefont {Zhang}},
  \bibinfo {author} {\bibfnamefont {T.~M.}\ \bibnamefont {Garoni}}, \bibinfo
  {author} {\bibfnamefont {A.~D.}\ \bibnamefont {Sokal}},\ and\ \bibinfo
  {author} {\bibfnamefont {A.}~\bibnamefont {Sportiello}},\ }\bibfield  {title}
  {\bibinfo {title} {Some geometric critical exponents for percolation and the
  random-cluster model},\ }\href {https://doi.org/10.1103/PhysRevE.81.020102}
  {\bibfield  {journal} {\bibinfo  {journal} {Phys. Rev. E}\ }\textbf {\bibinfo
  {volume} {81}},\ \bibinfo {pages} {020102} (\bibinfo {year}
  {2010})}\BibitemShut {NoStop}%
\bibitem [{\citenamefont {Cohen}\ \emph {et~al.}(2002)\citenamefont {Cohen},
  \citenamefont {ben Avraham},\ and\ \citenamefont {Havlin}}]{Cohen2002}%
  \BibitemOpen
  \bibfield  {author} {\bibinfo {author} {\bibfnamefont {R.}~\bibnamefont
  {Cohen}}, \bibinfo {author} {\bibfnamefont {D.}~\bibnamefont {ben Avraham}},\
  and\ \bibinfo {author} {\bibfnamefont {S.}~\bibnamefont {Havlin}},\
  }\bibfield  {title} {\bibinfo {title} {Percolation critical exponents in
  scale-free networks},\ }\href {https://doi.org/10.1103/physreve.66.036113}
  {\bibfield  {journal} {\bibinfo  {journal} {Phys. Rev. E}\ }\textbf {\bibinfo
  {volume} {66}},\ \bibinfo {pages} {036113} (\bibinfo {year}
  {2002})}\BibitemShut {NoStop}%
\bibitem [{\citenamefont {Lee}\ \emph {et~al.}(2004)\citenamefont {Lee},
  \citenamefont {Goh}, \citenamefont {Kahng},\ and\ \citenamefont
  {Kim}}]{Lee2004}%
  \BibitemOpen
  \bibfield  {author} {\bibinfo {author} {\bibfnamefont {D.-S.}\ \bibnamefont
  {Lee}}, \bibinfo {author} {\bibfnamefont {K.-I.}\ \bibnamefont {Goh}},
  \bibinfo {author} {\bibfnamefont {B.}~\bibnamefont {Kahng}},\ and\ \bibinfo
  {author} {\bibfnamefont {D.}~\bibnamefont {Kim}},\ }\bibfield  {title}
  {\bibinfo {title} {Evolution of scale-free random graphs: Potts model
  formulation},\ }\href
  {https://doi.org/https://doi.org/10.1016/j.nuclphysb.2004.06.029} {\bibfield
  {journal} {\bibinfo  {journal} {Nucl. Phys. B}\ }\textbf {\bibinfo {volume}
  {696}},\ \bibinfo {pages} {351 } (\bibinfo {year} {2004})}\BibitemShut
  {NoStop}%
\bibitem [{\citenamefont {Cirigliano}\ \emph {et~al.}(2024)\citenamefont
  {Cirigliano}, \citenamefont {Timár},\ and\ \citenamefont
  {Castellano}}]{Cirigliano2024}%
  \BibitemOpen
  \bibfield  {author} {\bibinfo {author} {\bibfnamefont {L.}~\bibnamefont
  {Cirigliano}}, \bibinfo {author} {\bibfnamefont {G.}~\bibnamefont {Timár}},\
  and\ \bibinfo {author} {\bibfnamefont {C.}~\bibnamefont {Castellano}},\
  }\bibfield  {title} {\bibinfo {title} {Scaling and universality for
  percolation in random networks: A unified view},\ }\href
  {https://doi.org/10.1103/physreve.110.064303} {\bibfield  {journal} {\bibinfo
   {journal} {Phys. Rev. E}\ }\textbf {\bibinfo {volume} {110}},\ \bibinfo
  {pages} {064303} (\bibinfo {year} {2024})}\BibitemShut {NoStop}%
\bibitem [{\citenamefont {Wu}\ \emph {et~al.}(2007)\citenamefont {Wu},
  \citenamefont {Lagorio}, \citenamefont {Braunstein}, \citenamefont {Cohen},
  \citenamefont {Havlin},\ and\ \citenamefont {Stanley}}]{Wu2007}%
  \BibitemOpen
  \bibfield  {author} {\bibinfo {author} {\bibfnamefont {Z.}~\bibnamefont
  {Wu}}, \bibinfo {author} {\bibfnamefont {C.}~\bibnamefont {Lagorio}},
  \bibinfo {author} {\bibfnamefont {L.~A.}\ \bibnamefont {Braunstein}},
  \bibinfo {author} {\bibfnamefont {R.}~\bibnamefont {Cohen}}, \bibinfo
  {author} {\bibfnamefont {S.}~\bibnamefont {Havlin}},\ and\ \bibinfo {author}
  {\bibfnamefont {H.~E.}\ \bibnamefont {Stanley}},\ }\bibfield  {title}
  {\bibinfo {title} {Numerical evaluation of the upper critical dimension of
  percolation in scale-free networks},\ }\href
  {https://doi.org/10.1103/PhysRevE.75.066110} {\bibfield  {journal} {\bibinfo
  {journal} {Phys. Rev. E}\ }\textbf {\bibinfo {volume} {75}},\ \bibinfo
  {pages} {066110} (\bibinfo {year} {2007})}\BibitemShut {NoStop}%
\bibitem [{\citenamefont {Zhao}\ \emph {et~al.}(2025)\citenamefont {Zhao},
  \citenamefont {Yang}, \citenamefont {Peng}, \citenamefont {Liu},\ and\
  \citenamefont {Li}}]{Zhao2025}%
  \BibitemOpen
  \bibfield  {author} {\bibinfo {author} {\bibfnamefont {X.}~\bibnamefont
  {Zhao}}, \bibinfo {author} {\bibfnamefont {L.}~\bibnamefont {Yang}}, \bibinfo
  {author} {\bibfnamefont {D.}~\bibnamefont {Peng}}, \bibinfo {author}
  {\bibfnamefont {R.-R.}\ \bibnamefont {Liu}},\ and\ \bibinfo {author}
  {\bibfnamefont {M.}~\bibnamefont {Li}},\ }\bibfield  {title} {\bibinfo
  {title} {Finite-size scaling of percolation on scale-free networks},\ }\href
  {https://doi.org/10.1016/j.chaos.2025.117076} {\bibfield  {journal} {\bibinfo
   {journal} {Chaos Soliton Fract.}\ }\textbf {\bibinfo {volume} {200}},\
  \bibinfo {pages} {117076} (\bibinfo {year} {2025})}\BibitemShut {NoStop}%
\bibitem [{\citenamefont {Molloy}\ and\ \citenamefont
  {Reed}(1995)}]{Molloy1995}%
  \BibitemOpen
  \bibfield  {author} {\bibinfo {author} {\bibfnamefont {M.}~\bibnamefont
  {Molloy}}\ and\ \bibinfo {author} {\bibfnamefont {B.}~\bibnamefont {Reed}},\
  }\bibfield  {title} {\bibinfo {title} {A critical point for random graphs
  with a given degree sequence},\ }\href
  {https://doi.org/10.1002/rsa.3240060204} {\bibfield  {journal} {\bibinfo
  {journal} {Random Struct. Algor.}\ }\textbf {\bibinfo {volume} {6}},\
  \bibinfo {pages} {161} (\bibinfo {year} {1995})}\BibitemShut {NoStop}%
\bibitem [{\citenamefont {Newman}\ \emph {et~al.}(2001)\citenamefont {Newman},
  \citenamefont {Strogatz},\ and\ \citenamefont {Watts}}]{Newman2001}%
  \BibitemOpen
  \bibfield  {author} {\bibinfo {author} {\bibfnamefont {M.~E.~J.}\
  \bibnamefont {Newman}}, \bibinfo {author} {\bibfnamefont {S.~H.}\
  \bibnamefont {Strogatz}},\ and\ \bibinfo {author} {\bibfnamefont {D.~J.}\
  \bibnamefont {Watts}},\ }\bibfield  {title} {\bibinfo {title} {Random graphs
  with arbitrary degree distributions and their applications},\ }\href
  {https://doi.org/10.1103/physreve.64.026118} {\bibfield  {journal} {\bibinfo
  {journal} {Phys. Rev. E}\ }\textbf {\bibinfo {volume} {64}},\ \bibinfo
  {pages} {026118} (\bibinfo {year} {2001})}\BibitemShut {NoStop}%
\bibitem [{\citenamefont {D'Souza}\ \emph {et~al.}(2019)\citenamefont
  {D'Souza}, \citenamefont {G\'{o}mez-Garde\~{n}es}, \citenamefont {Nagler},\
  and\ \citenamefont {Arenas}}]{D’Souza2019}%
  \BibitemOpen
  \bibfield  {author} {\bibinfo {author} {\bibfnamefont {R.~M.}\ \bibnamefont
  {D'Souza}}, \bibinfo {author} {\bibfnamefont {J.}~\bibnamefont
  {G\'{o}mez-Garde\~{n}es}}, \bibinfo {author} {\bibfnamefont {J.}~\bibnamefont
  {Nagler}},\ and\ \bibinfo {author} {\bibfnamefont {A.}~\bibnamefont
  {Arenas}},\ }\bibfield  {title} {\bibinfo {title} {Explosive phenomena in
  complex networks},\ }\href {https://doi.org/10.1080/00018732.2019.1650450}
  {\bibfield  {journal} {\bibinfo  {journal} {Adv. Phys.}\ }\textbf {\bibinfo
  {volume} {68}},\ \bibinfo {pages} {123} (\bibinfo {year} {2019})}\BibitemShut
  {NoStop}%
\bibitem [{\citenamefont {Thorpe}\ \emph {et~al.}(2000)\citenamefont {Thorpe},
  \citenamefont {Jacobs}, \citenamefont {Chubynsky},\ and\ \citenamefont
  {Phillips}}]{Thorpe2000}%
  \BibitemOpen
  \bibfield  {author} {\bibinfo {author} {\bibfnamefont {M.~F.}\ \bibnamefont
  {Thorpe}}, \bibinfo {author} {\bibfnamefont {D.~J.}\ \bibnamefont {Jacobs}},
  \bibinfo {author} {\bibfnamefont {M.~V.}\ \bibnamefont {Chubynsky}},\ and\
  \bibinfo {author} {\bibfnamefont {J.~C.}\ \bibnamefont {Phillips}},\
  }\bibfield  {title} {\bibinfo {title} {Self-organization in network
  glasses},\ }\href {https://doi.org/10.1016/s0022-3093(99)00856-x} {\bibfield
  {journal} {\bibinfo  {journal} {J. Non-Cryst. Solids}\ }\textbf {\bibinfo
  {volume} {266–269}},\ \bibinfo {pages} {859} (\bibinfo {year}
  {2000})}\BibitemShut {NoStop}%
\bibitem [{\citenamefont {Broedersz}\ \emph {et~al.}(2011)\citenamefont
  {Broedersz}, \citenamefont {Mao}, \citenamefont {Lubensky},\ and\
  \citenamefont {MacKintosh}}]{Broedersz2011}%
  \BibitemOpen
  \bibfield  {author} {\bibinfo {author} {\bibfnamefont {C.~P.}\ \bibnamefont
  {Broedersz}}, \bibinfo {author} {\bibfnamefont {X.}~\bibnamefont {Mao}},
  \bibinfo {author} {\bibfnamefont {T.~C.}\ \bibnamefont {Lubensky}},\ and\
  \bibinfo {author} {\bibfnamefont {F.~C.}\ \bibnamefont {MacKintosh}},\
  }\bibfield  {title} {\bibinfo {title} {Criticality and isostaticity in fibre
  networks},\ }\href {https://doi.org/10.1038/nphys2127} {\bibfield  {journal}
  {\bibinfo  {journal} {Nat. Phys.}\ }\textbf {\bibinfo {volume} {7}},\
  \bibinfo {pages} {983} (\bibinfo {year} {2011})}\BibitemShut {NoStop}%
\bibitem [{\citenamefont {Zhang}\ \emph {et~al.}(2019)\citenamefont {Zhang},
  \citenamefont {Zhang}, \citenamefont {Bouzid}, \citenamefont {Rocklin},
  \citenamefont {Del~Gado},\ and\ \citenamefont {Mao}}]{Zhang2019}%
  \BibitemOpen
  \bibfield  {author} {\bibinfo {author} {\bibfnamefont {S.}~\bibnamefont
  {Zhang}}, \bibinfo {author} {\bibfnamefont {L.}~\bibnamefont {Zhang}},
  \bibinfo {author} {\bibfnamefont {M.}~\bibnamefont {Bouzid}}, \bibinfo
  {author} {\bibfnamefont {D.~Z.}\ \bibnamefont {Rocklin}}, \bibinfo {author}
  {\bibfnamefont {E.}~\bibnamefont {Del~Gado}},\ and\ \bibinfo {author}
  {\bibfnamefont {X.}~\bibnamefont {Mao}},\ }\bibfield  {title} {\bibinfo
  {title} {Correlated rigidity percolation and colloidal gels},\ }\href
  {https://doi.org/10.1103/physrevlett.123.058001} {\bibfield  {journal}
  {\bibinfo  {journal} {Phys. Rev. Lett.}\ }\textbf {\bibinfo {volume} {123}},\
  \bibinfo {pages} {058001} (\bibinfo {year} {2019})}\BibitemShut {NoStop}%
\bibitem [{\citenamefont {Rouwhorst}\ \emph
  {et~al.}(2020{\natexlab{a}})\citenamefont {Rouwhorst}, \citenamefont {Ness},
  \citenamefont {Stoyanov}, \citenamefont {Zaccone},\ and\ \citenamefont
  {Schall}}]{Rouwhorst2020}%
  \BibitemOpen
  \bibfield  {author} {\bibinfo {author} {\bibfnamefont {J.}~\bibnamefont
  {Rouwhorst}}, \bibinfo {author} {\bibfnamefont {C.}~\bibnamefont {Ness}},
  \bibinfo {author} {\bibfnamefont {S.}~\bibnamefont {Stoyanov}}, \bibinfo
  {author} {\bibfnamefont {A.}~\bibnamefont {Zaccone}},\ and\ \bibinfo {author}
  {\bibfnamefont {P.}~\bibnamefont {Schall}},\ }\bibfield  {title} {\bibinfo
  {title} {Nonequilibrium continuous phase transition in colloidal gelation
  with short-range attraction},\ }\href
  {https://doi.org/10.1038/s41467-020-17353-8} {\bibfield  {journal} {\bibinfo
  {journal} {Nat. Commun.}\ }\textbf {\bibinfo {volume} {11}},\ \bibinfo
  {pages} {3558} (\bibinfo {year} {2020}{\natexlab{a}})}\BibitemShut {NoStop}%
\bibitem [{\citenamefont {Rouwhorst}\ \emph
  {et~al.}(2020{\natexlab{b}})\citenamefont {Rouwhorst}, \citenamefont
  {Schall}, \citenamefont {Ness}, \citenamefont {Blijdenstein},\ and\
  \citenamefont {Zaccone}}]{Rouwhorst2020a}%
  \BibitemOpen
  \bibfield  {author} {\bibinfo {author} {\bibfnamefont {J.}~\bibnamefont
  {Rouwhorst}}, \bibinfo {author} {\bibfnamefont {P.}~\bibnamefont {Schall}},
  \bibinfo {author} {\bibfnamefont {C.}~\bibnamefont {Ness}}, \bibinfo {author}
  {\bibfnamefont {T.}~\bibnamefont {Blijdenstein}},\ and\ \bibinfo {author}
  {\bibfnamefont {A.}~\bibnamefont {Zaccone}},\ }\bibfield  {title} {\bibinfo
  {title} {Nonequilibrium master kinetic equation modeling of colloidal
  gelation},\ }\href {https://doi.org/10.1103/physreve.102.022602} {\bibfield
  {journal} {\bibinfo  {journal} {Phys. Rev. E}\ }\textbf {\bibinfo {volume}
  {102}},\ \bibinfo {pages} {022602} (\bibinfo {year}
  {2020}{\natexlab{b}})}\BibitemShut {NoStop}%
\bibitem [{\citenamefont {Zaccone}\ and\ \citenamefont
  {Scossa-Romano}(2011)}]{Zaccone2011}%
  \BibitemOpen
  \bibfield  {author} {\bibinfo {author} {\bibfnamefont {A.}~\bibnamefont
  {Zaccone}}\ and\ \bibinfo {author} {\bibfnamefont {E.}~\bibnamefont
  {Scossa-Romano}},\ }\bibfield  {title} {\bibinfo {title} {Approximate
  analytical description of the nonaffine response of amorphous solids},\
  }\href {https://doi.org/10.1103/physrevb.83.184205} {\bibfield  {journal}
  {\bibinfo  {journal} {Phys. Rev. B}\ }\textbf {\bibinfo {volume} {83}},\
  \bibinfo {pages} {184205} (\bibinfo {year} {2011})}\BibitemShut {NoStop}%
\bibitem [{\citenamefont {Henkes}\ \emph {et~al.}(2016)\citenamefont {Henkes},
  \citenamefont {Quint}, \citenamefont {Fily},\ and\ \citenamefont
  {Schwarz}}]{Henkes2016}%
  \BibitemOpen
  \bibfield  {author} {\bibinfo {author} {\bibfnamefont {S.}~\bibnamefont
  {Henkes}}, \bibinfo {author} {\bibfnamefont {D.~A.}\ \bibnamefont {Quint}},
  \bibinfo {author} {\bibfnamefont {Y.}~\bibnamefont {Fily}},\ and\ \bibinfo
  {author} {\bibfnamefont {J.}~\bibnamefont {Schwarz}},\ }\bibfield  {title}
  {\bibinfo {title} {Rigid cluster decomposition reveals criticality in
  frictional jamming},\ }\href {https://doi.org/10.1103/physrevlett.116.028301}
  {\bibfield  {journal} {\bibinfo  {journal} {Phys. Rev. Lett.}\ }\textbf
  {\bibinfo {volume} {116}},\ \bibinfo {pages} {028301} (\bibinfo {year}
  {2016})}\BibitemShut {NoStop}%
\bibitem [{\citenamefont {Dashti}\ \emph {et~al.}(2023)\citenamefont {Dashti},
  \citenamefont {Saberi}, \citenamefont {Rahbari},\ and\ \citenamefont
  {Kurths}}]{Dashti2023}%
  \BibitemOpen
  \bibfield  {author} {\bibinfo {author} {\bibfnamefont {H.}~\bibnamefont
  {Dashti}}, \bibinfo {author} {\bibfnamefont {A.~A.}\ \bibnamefont {Saberi}},
  \bibinfo {author} {\bibfnamefont {S.}~\bibnamefont {Rahbari}},\ and\ \bibinfo
  {author} {\bibfnamefont {J.}~\bibnamefont {Kurths}},\ }\bibfield  {title}
  {\bibinfo {title} {Emergence of rigidity percolation in flowing granular
  systems},\ }\href {https://doi.org/10.1126/sciadv.adh5586} {\bibfield
  {journal} {\bibinfo  {journal} {Sci. Adv.}\ }\textbf {\bibinfo {volume}
  {9}},\ \bibinfo {pages} {eadh5586} (\bibinfo {year} {2023})}\BibitemShut
  {NoStop}%
\bibitem [{\citenamefont {Vinutha}\ \emph {et~al.}(2023)\citenamefont
  {Vinutha}, \citenamefont {Diaz~Ruiz}, \citenamefont {Mao}, \citenamefont
  {Chakraborty},\ and\ \citenamefont {Del~Gado}}]{Vinutha2023}%
  \BibitemOpen
  \bibfield  {author} {\bibinfo {author} {\bibfnamefont {H.~A.}\ \bibnamefont
  {Vinutha}}, \bibinfo {author} {\bibfnamefont {F.~D.}\ \bibnamefont
  {Diaz~Ruiz}}, \bibinfo {author} {\bibfnamefont {X.}~\bibnamefont {Mao}},
  \bibinfo {author} {\bibfnamefont {B.}~\bibnamefont {Chakraborty}},\ and\
  \bibinfo {author} {\bibfnamefont {E.}~\bibnamefont {Del~Gado}},\ }\bibfield
  {title} {\bibinfo {title} {Stress–stress correlations reveal force chains
  in gels},\ }\href {https://doi.org/10.1063/5.0131473} {\bibfield  {journal}
  {\bibinfo  {journal} {J. Chem. Phys.}\ }\textbf {\bibinfo {volume} {158}},\
  \bibinfo {pages} {114104} (\bibinfo {year} {2023})}\BibitemShut {NoStop}%
\bibitem [{\citenamefont {Petridou}\ \emph {et~al.}(2021)\citenamefont
  {Petridou}, \citenamefont {Corominas-Murtra}, \citenamefont {Heisenberg},\
  and\ \citenamefont {Hannezo}}]{Petridou2021}%
  \BibitemOpen
  \bibfield  {author} {\bibinfo {author} {\bibfnamefont {N.~I.}\ \bibnamefont
  {Petridou}}, \bibinfo {author} {\bibfnamefont {B.}~\bibnamefont
  {Corominas-Murtra}}, \bibinfo {author} {\bibfnamefont {C.-P.}\ \bibnamefont
  {Heisenberg}},\ and\ \bibinfo {author} {\bibfnamefont {E.}~\bibnamefont
  {Hannezo}},\ }\bibfield  {title} {\bibinfo {title} {Rigidity percolation
  uncovers a structural basis for embryonic tissue phase transitions},\ }\href
  {https://doi.org/10.1016/j.cell.2021.02.017} {\bibfield  {journal} {\bibinfo
  {journal} {Cell}\ }\textbf {\bibinfo {volume} {184}},\ \bibinfo {pages}
  {1914} (\bibinfo {year} {2021})}\BibitemShut {NoStop}%
\bibitem [{\citenamefont {Rozman}\ \emph {et~al.}(2024)\citenamefont {Rozman},
  \citenamefont {Krajnc},\ and\ \citenamefont {Ziherl}}]{Rozman2024}%
  \BibitemOpen
  \bibfield  {author} {\bibinfo {author} {\bibfnamefont {J.}~\bibnamefont
  {Rozman}}, \bibinfo {author} {\bibfnamefont {M.}~\bibnamefont {Krajnc}},\
  and\ \bibinfo {author} {\bibfnamefont {P.}~\bibnamefont {Ziherl}},\
  }\bibfield  {title} {\bibinfo {title} {Basolateral mechanics prevents
  rigidity transition in epithelial monolayers},\ }\href
  {https://doi.org/10.1103/PhysRevLett.133.168401} {\bibfield  {journal}
  {\bibinfo  {journal} {Phys. Rev. Lett.}\ }\textbf {\bibinfo {volume} {133}},\
  \bibinfo {pages} {168401} (\bibinfo {year} {2024})}\BibitemShut {NoStop}%
\bibitem [{\citenamefont {Lübeck}\ and\ \citenamefont
  {Usadel}(1997)}]{Luebeck1997}%
  \BibitemOpen
  \bibfield  {author} {\bibinfo {author} {\bibfnamefont {S.}~\bibnamefont
  {Lübeck}}\ and\ \bibinfo {author} {\bibfnamefont {K.~D.}\ \bibnamefont
  {Usadel}},\ }\bibfield  {title} {\bibinfo {title} {Numerical determination of
  the avalanche exponents of the {B}ak-{T}ang-{W}iesenfeld model},\ }\href
  {https://doi.org/10.1103/physreve.55.4095} {\bibfield  {journal} {\bibinfo
  {journal} {Phys. Rev. E}\ }\textbf {\bibinfo {volume} {55}},\ \bibinfo
  {pages} {4095} (\bibinfo {year} {1997})}\BibitemShut {NoStop}%
\bibitem [{\citenamefont {Kadanoff}\ \emph {et~al.}(1989)\citenamefont
  {Kadanoff}, \citenamefont {Nagel}, \citenamefont {Wu},\ and\ \citenamefont
  {Zhou}}]{Kadanoff1989}%
  \BibitemOpen
  \bibfield  {author} {\bibinfo {author} {\bibfnamefont {L.~P.}\ \bibnamefont
  {Kadanoff}}, \bibinfo {author} {\bibfnamefont {S.~R.}\ \bibnamefont {Nagel}},
  \bibinfo {author} {\bibfnamefont {L.}~\bibnamefont {Wu}},\ and\ \bibinfo
  {author} {\bibfnamefont {S.-M.}\ \bibnamefont {Zhou}},\ }\bibfield  {title}
  {\bibinfo {title} {Scaling and universality in avalanches},\ }\href
  {https://doi.org/10.1103/physreva.39.6524} {\bibfield  {journal} {\bibinfo
  {journal} {Phys. Rev. A}\ }\textbf {\bibinfo {volume} {39}},\ \bibinfo
  {pages} {6524} (\bibinfo {year} {1989})}\BibitemShut {NoStop}%
\bibitem [{\citenamefont {Manna}(1991)}]{Manna1991}%
  \BibitemOpen
  \bibfield  {author} {\bibinfo {author} {\bibfnamefont {S.}~\bibnamefont
  {Manna}},\ }\bibfield  {title} {\bibinfo {title} {Critical exponents of the
  sandpile models in two dimensions},\ }\href
  {https://doi.org/10.1016/0378-4371(91)90063-i} {\bibfield  {journal}
  {\bibinfo  {journal} {Physica A}\ }\textbf {\bibinfo {volume} {179}},\
  \bibinfo {pages} {249} (\bibinfo {year} {1991})}\BibitemShut {NoStop}%
\bibitem [{\citenamefont {Milshtein}\ \emph {et~al.}(1998)\citenamefont
  {Milshtein}, \citenamefont {Biham},\ and\ \citenamefont
  {Solomon}}]{Milshtein1998}%
  \BibitemOpen
  \bibfield  {author} {\bibinfo {author} {\bibfnamefont {E.}~\bibnamefont
  {Milshtein}}, \bibinfo {author} {\bibfnamefont {O.}~\bibnamefont {Biham}},\
  and\ \bibinfo {author} {\bibfnamefont {S.}~\bibnamefont {Solomon}},\
  }\bibfield  {title} {\bibinfo {title} {Universality classes in isotropic,
  {A}belian, and non-{A}belian sandpile models},\ }\href
  {https://doi.org/10.1103/physreve.58.303} {\bibfield  {journal} {\bibinfo
  {journal} {Phys. Rev. E}\ }\textbf {\bibinfo {volume} {58}},\ \bibinfo
  {pages} {303} (\bibinfo {year} {1998})}\BibitemShut {NoStop}%
\bibitem [{\citenamefont {Liu}\ \emph {et~al.}(2025)\citenamefont {Liu},
  \citenamefont {Xiao}, \citenamefont {Fan},\ and\ \citenamefont
  {Deng}}]{Liu2025}%
  \BibitemOpen
  \bibfield  {author} {\bibinfo {author} {\bibfnamefont {Z.}~\bibnamefont
  {Liu}}, \bibinfo {author} {\bibfnamefont {T.}~\bibnamefont {Xiao}}, \bibinfo
  {author} {\bibfnamefont {Z.}~\bibnamefont {Fan}},\ and\ \bibinfo {author}
  {\bibfnamefont {Y.}~\bibnamefont {Deng}},\ }\bibfield  {title} {\bibinfo
  {title} {Two-dimensional percolation model with long-range interaction},\
  }\href {https://doi.org/10.48550/ARXIV.2509.18035} {\bibfield  {journal}
  {\bibinfo  {journal} {arXiv preprint}\ ,\ \bibinfo {pages} {2509.18035}}
  (\bibinfo {year} {2025})}\BibitemShut {NoStop}%
\bibitem [{\citenamefont {Yang}\ and\ \citenamefont {Li}(2024)}]{Yang2024}%
  \BibitemOpen
  \bibfield  {author} {\bibinfo {author} {\bibfnamefont {L.}~\bibnamefont
  {Yang}}\ and\ \bibinfo {author} {\bibfnamefont {M.}~\bibnamefont {Li}},\
  }\bibfield  {title} {\bibinfo {title} {Emergence of biconnected clusters in
  explosive percolation},\ }\href {https://doi.org/10.1103/physreve.110.014122}
  {\bibfield  {journal} {\bibinfo  {journal} {Phys. Rev. E}\ }\textbf {\bibinfo
  {volume} {110}},\ \bibinfo {pages} {014122} (\bibinfo {year}
  {2024})}\BibitemShut {NoStop}%
\bibitem [{\citenamefont {Fang}\ \emph {et~al.}(2024)\citenamefont {Fang},
  \citenamefont {Lin}, \citenamefont {Meng}, \citenamefont {Chen},
  \citenamefont {Nagler}, \citenamefont {Deng},\ and\ \citenamefont
  {Fan}}]{Fang2024}%
  \BibitemOpen
  \bibfield  {author} {\bibinfo {author} {\bibfnamefont {S.}~\bibnamefont
  {Fang}}, \bibinfo {author} {\bibfnamefont {Q.}~\bibnamefont {Lin}}, \bibinfo
  {author} {\bibfnamefont {J.}~\bibnamefont {Meng}}, \bibinfo {author}
  {\bibfnamefont {B.}~\bibnamefont {Chen}}, \bibinfo {author} {\bibfnamefont
  {J.}~\bibnamefont {Nagler}}, \bibinfo {author} {\bibfnamefont
  {Y.}~\bibnamefont {Deng}},\ and\ \bibinfo {author} {\bibfnamefont
  {J.}~\bibnamefont {Fan}},\ }\bibfield  {title} {\bibinfo {title} {Universal
  scaling of gap dynamics in percolation},\ }\href
  {https://arxiv.org/abs/2410.24068} {\bibfield  {journal} {\bibinfo  {journal}
  {arXiv preprint}\ ,\ \bibinfo {pages} {2410.24068}} (\bibinfo {year}
  {2024})}\BibitemShut {NoStop}%
\end{thebibliography}%

\end{document}